\documentclass[num-refs]{wiley-article}

\usepackage{siunitx}
\usepackage{subfig}
\usepackage{color}
\usepackage{longtable}
\usepackage{algorithm}%
\usepackage{algorithmicx}%
\usepackage{algpseudocode}%

\usepackage{caption}
\usepackage{multirow}
\usepackage{enumitem}
\usepackage{xcolor}

\papertype{Original Article}
\paperfield{Journal Section}

\usepackage{fancyhdr}
\title {Exploring Gene Regulatory Interaction Networks and predicting therapeutic molecules for Hypopharyngeal Cancer and EGFR-mutated lung adenocarcinoma}

\author[1]{Abanti Bhattacharjya}
\author[1]{Md Manowarul Islam}
\author[2]{Md Ashraf Uddin}
\author[3]{Md. Alamin Talukder}

\author[4]{AKM Azad}
\author[2]{Sunil Aryal}

\author[5]{Bikash Kumar Paul}
\author[5]{Wahia Tasnim}

\author[5]{Muhammad Ali Abdulllah Almoyad} 
\author[6]{Mohammad Ali Moni}

\affil[1]{Department of Computer Science and Engineering, Jagannath University, Dhaka, Bangladesh}
\affil[2]{School of Information Technology, Deakin University, Waurn Ponds Campus, Geelong, Australia}
\affil[3]{Department of Computer Science and Engineering, International University of Business Agriculture and Technology, Dhaka, Bangladesh}

\affil[4]{Department of Mathematics and Statistics, Faculty of Science, Imam Mohammad Ibn Saud Islamic University (IMSIU), Riyadh 13318, Saudi Arabia}
\affil[5]{Department of Information and Communication Technology, Mawlana Bhashani Science and Technology University, Bangladesh}
\affil[5]{Department of Software Engineering, Daffodil International University,  Dhaka, Bangladesh}
\affil[5]{Department of Basic Medical Sciences, College of Applied Medical Sciences, King Khalid University, Saudi Arabia}
\affil[6]{Artificial Intelligence \& Data Science, Faculty of Health and Behavioural Sciences, The University of Queensland, Australia}

\corraddress{Mohammad Ali Moni and Md Manowarul Islam}
\corremail{m.moni@uq.edu.au and manowar@cse.jnu.ac.bd}

\runningauthor{Bhattacharjya, Abanti et al.}

\begin{document}

\begin{frontmatter}
\maketitle

\begin{abstract}
With the advent of Information technology, the Bioinformatics research field is becoming increasingly attractive to researchers and academicians. The recent development of various Bioinformatics toolkits has facilitated the rapid processing and analysis of vast quantities of biological data for human perception. Most studies focus on locating two connected diseases and making some observations to construct diverse gene regulatory interaction networks, a forerunner to general drug design for curing illness. For instance, Hypopharyngeal cancer is a disease that is associated with EGFR-mutated lung adenocarcinoma. In this study, we select EGFR-mutated lung adenocarcinoma and Hypopharyngeal cancer by finding the Lung metastases in hypopharyngeal cancer.

To conduct this study, we collect Mircorarray datasets from GEO (Gene Expression Omnibus), an online database controlled by NCBI. Differentially expressed genes, common genes, and hub genes between the selected two diseases are detected for the succeeding move. Our research findings have suggested common therapeutic molecules for the selected diseases based on 10 hub genes with the highest interactions according to the degree topology method and the maximum clique centrality (MCC).  Our suggested therapeutic molecules will be fruitful for patients with those two diseases simultaneously.


\end{abstract}
\end{frontmatter}

\section{Introduction}
\label{intro}

Bioinformatics, which combines the capabilities of computer science with biology, has expanded significantly in recent years\cite{phoebe2008identifying}. Several Bioinformatics toolkits are leveraged to achieve the desired result for the experiment. Bioinformatics can research the molecular causes of sickness, describe the disease's situation from the gene's nook, and reduce the amount of time and money spent on the process by utilizing computer abilities to narrow the scope of the investigation and improve the quality of the results \cite{phoebe2008identifying}. 
200 different cell types and 100 different cancers have been found among the 100 trillion cells in the human body \cite{hanif2018head}. Cancer is a group of diseases characterized by abnormal cell proliferation that attacks like a crab and utilizes its numerous claws to try to kill its target \cite{hanif2018head, talukder2023efficient, talukder2023empowering, khatun2023cancer, talukder2024mlstl, talukder2024machine, talukder2024securing}.


The prognosis of tumors that originate from other head \& neck sites is better than that of Hypopharyngeal cancer, a less frequent type of head \& neck cancer \cite{sanders2022hypopharyngeal}. With only 15–30\% of patients living for more than five years, Hypopharyngeal carcinoma, which makes up around 5\% of all head \& neck malignancies, has a horrible prognosis \cite{hoffmann2005human} \cite{rodrigo2015prevalence}. Two common risk factors for Hypopharyngeal carcinoma, are alcohol consumption and smoking \cite{lindquist2007human}. According to the American Cancer Society, the Human Papillomavirus also causes Hypopharyngeal carcinoma. 

The epidermal growth factor receptor gene is the most commonly mutated gene in lung cancer (EGFR) \cite{jin2021egfr}. Lung squamous cell carcinoma (SCC), which has an estimated frequency of 3\% to 18\%, is comparatively uncommon compared to lung adenocarcinoma (1–10) \cite{jin2021egfr}. Lung cancer, which comprises both small and non-small cells, is the leading cause of cancer-related death globally \cite{bray2018global, talukder2022machine}. The world's highest incidence and fatality rates are associated with the most prevalent kinds of cancer \cite{wild2020world}. Risk factors for lung cancer include smoking, passive smoking, age, gender, family history, chronic lung disease, chest radiotherapy, diet, obesity, physical activity, alcohol consumption, employment, education, and income \cite{mukti2014score}. The Human Papillomavirus might potentially increase the risk of developing Lung Cancer \cite{klein2009incidence}.

Head and Neck Cancers as well as Lung Cancers pose significant challenges to global health \cite{van2016differentiating}. Head and Neck Cancer is among the most frequent malignancies to migrate to the lungs \cite{kaifi2010indications} \cite{pfannschmidt2012surgical} \cite{kondo2005surgical}. Following bone and soft tissue sarcomas, \cite{saleh2018surgical} recognized head-neck cancer as the third most frequent reason for pulmonary mastectomy. \cite{ferlay2015cancer} reported in 2012 that there were 686,000 new instances of head and neck cancer, 1,825,000 new cases of lung cancer, and a combined mortality rate of 5\% and 19\%, respectively. As Hypopharyngeal cancer is a type of head and neck cancer and EGFR-mutated lung adenocarcinoma is also one form of Lung cancer. So, we can claim that patients with EGFR-mutated lung adenocarcinoma may have the potential to develop Hypopharyngeal cancer, according to the preceding statistic. Also, Hypopharyngeal cancer may potentially spread to lung adenocarcinoma with EGFR mutation, and lung adenocarcinoma with EGFR mutation may potentially metastasize to Hypopharyngeal cancer.  \cite{shen2021lung} This paper's researchers did an analysis of the general population's lung metastases in newly diagnosed hypopharyngeal cancer. \cite{cancerca} According to the Canadian Cancer Society, lung cancer may develop if hypopharyngeal cancer progresses.  Therefore, this concludes that they are related genetically because they share genes. This set of shared genes is restrained by regulatory interaction network pathways.


In this research, we aim to look into common DEGs(Differentially Expressed Genes), Hub genes, Various Gene Regulatory networks, and Therapeutic Molecule for Hypopharyngeal cancer and EGFR-mutated lung adenocarcinoma using Bioinformatics technology.

\begin{figure}
     \centering
     \includegraphics[scale=0.45]{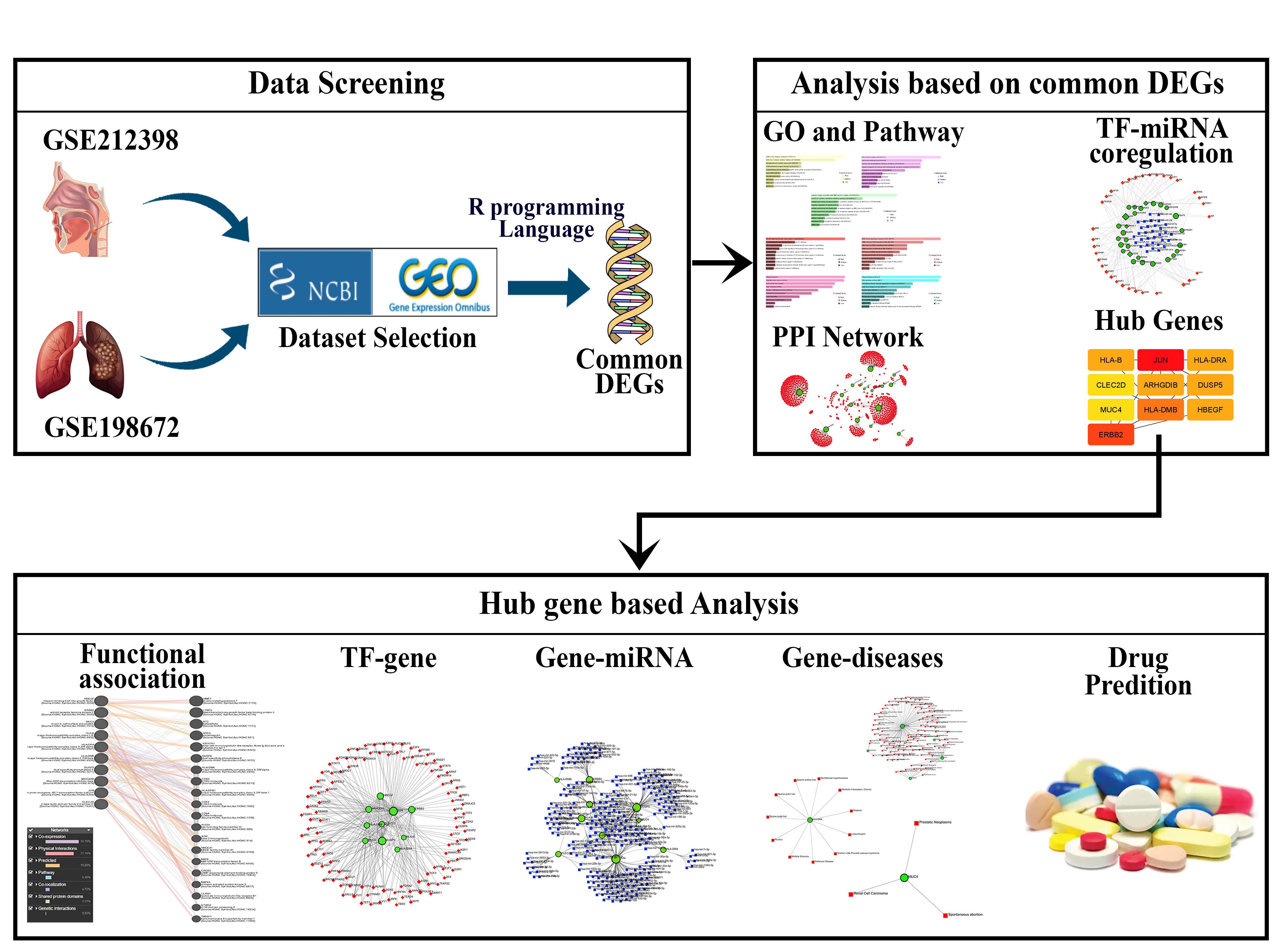}
     \caption{Diagram representing the proposed methodology of the current research. For Hypopharyngeal Cancer and EGFR-mutated Lung Adenocarcinoma, two datasets are used. Each dataset has eight samples. Using the R programming language, the DEGs (Differentially Expressed Genes) from those two datasets are retrieved. The VENNY tool is used to find out the common genes between these two diseases. With the aid of these widespread DEGs, GO terms, pathways, PPI networks, TF-miRNA, and Hub genes are identified. Functional association, TF-gene, Gene-miRNA, Gene-disease, and Some therapeutic compounds are anticipated based on the hub genes of individuals with Hypopharyngeal Cancer and EGFR-mutated Lung Adenocarcinoma who have these 2 diseases concurrently.}
     \label{fig: Flow Diagram}
 \end{figure}

We used two datasets for EGFR-mutated Lung Adenocarcinoma and Hypopharyngeal Cancer. Each of these datasets has eight samples. The DEGs (Differentially Expressed Genes) shared by these two datasets are extracted using the R programming language. These widely distributed DEGs help to identify GO terms, pathways, PPI networks, and TF-miRNA. Based on the hub genes of patients with Hypopharyngeal Cancer and EGFR-mutated Lung Adenocarcinoma who have these 2 diseases concurrently, certain therapeutic compounds are envisaged. Hypopharyngeal cancer and EGFR-mutated lung adenocarcinoma can be associated with each other directly or indirectly. These diseases have some common interrelated genes. Gene regulatory interaction networks are developed by using different types of Bioinformatics tools. The PPI network is visualized, and common drugs are developed for the selected two associated diseases. A PPI network describes the connections between proteins in a biological system in the context of biological study. The process of visualizing this network usually entails producing a map or graphical depiction that shows the connections and interactions between various proteins. By illustrating these relationships graphically, scientists can better understand the intricate biological mechanisms involving proteins and possibly pinpoint important hubs or nodes in the network that are essential to cellular operations or disease processes. Visualizing the PPI network is all things considered, a step towards better understanding the complexities of molecular interactions through the analysis of biological data. Developing common drugs indicates the creation of common medications for the selected diseases. In addition, designing one common drug for two associated diseases decreases the amount of drug one should absorb for the diseases separately. 

Microarray data exploration is among the most well-known techniques used for extensive investigations of gene expression, and high throughput technologies are becoming more and more important in the field of biomedical research \cite{sturn2002genesis}. Researchers in genetics can analyze gene expression simultaneously with the help of microarray studies \cite{lee2000importance}. This research attempts to discover the relationship between the selected 2 diseases. GSE212398 for Hypopharyngeal Cancer and GSE198672 for EGFR-mutated Lung Adenocarcinoma, datasets are used for the investigation. The NCBI's GEO database served as the source for both dataset selections. Shared DEGs are collected from those 2 datasets. Figure \ref{fig: Flow Diagram} presents the proposed methodology's flow diagram.

This work aims to identify targeted therapeutic molecules for these 2 diseases. Targeted therapy has very often a remarkable effect against cancer. Drug compounds can serve multiple purposes in cancer treatment, including reducing the size of tumors before surgery, removing any remaining cancer cells after surgery, or as a last resort when other treatments are ineffective or cancer recurs.


Our contributions are summarized as follows.

\begin{itemize}
    \item To propose a Bioinformatics framework for integratively analyzing expression profiles of Lung adenocarcinoma and Hypopharyngeal cancer samples to find commonly found biomarkers.
     \item To conduct detailed downstream analyses based on commonly found biomarkers.
 \item Finally, propose some therapeutic agents for those biomarkers via drug-target analyses.
    
\end{itemize}

The rest of the paper is organized as follows. We review the related literature in Section \ref{relatedworks}. In section \ref{sec:Method}, the methodology is presented. The result analysis is conducted in the Section before concluding the paper in Section.

\section{Related Literature}
\label{relatedworks}   In this section, we discuss some earlier works based on some different diseases using the various bioinformatics approaches. 

\cite{islam2022identification} Molecular biomarker identification to suggest therapeutic targets for the creation of medicines to treat esophageal cancer. The authors collected these GSE93756, GSE94012, GSE104958, and GSE143822 datasets from the National Center for Biotechnology Information’s (NCBI) Gene Expression Omnibus. Using the R Language Limma Package, DEGs were collected by applying the adjusted value ($<$ 0.05). After that, DEGs' GO and Pathway Enrichment Analysis was done. And PPI and Clustering Analysis were also done based on the DEGs.
\cite{taz2021network} discussed idiopathic pulmonary fibrosis (IPF) people who have SARS-CoV-2 infections are genetically more likely to develop IPF. The GEO NCBI database was used to collect the GSE147507 and GSE35145 datasets. DEGs for GSE35145 were retrieved using the GEO2R tool that is included by default with the dataset in GEO, while DEGs for GSE147507 were gathered using the R programming language. Adjusted P-value ($<$ 0.05) and log2-fold change (absolute) ($>$ 1.0) were used as the cut-off criteria. Common genes for these 2 diseases were used for gene set enrichment analysis, PPIs network construction, hub gene searching and module examination, TF-miRNA identification, and candidate drug suggestion. 
\cite{taz2021identification} Identification of the SARS-CoV-2 biomarkers and pathways that complicate the condition in patients with pulmonary arterial hypertension. The R programming language's limma and DESeq2 packages were used to gather DEGs of GSE147507 for SARS-CoV-2 infection in human lung epithelial cells. DEGs of GSE117261 for PAH lung were found through the GEO2R tool of the GEO NCBI database. Adjusted P-value ($<$ 0.05) and log2-fold change (absolute) ($>$ 1.0) were used as the cut-off criteria. For the objective of controlling the false discovery rate, the Benjamini-Hochberg approach was used on both datasets. Common DEGs from these 2 datasets were used for further analysis.

\cite{chen2019screening} Employing Bioinformatics analysis to screen for and identify potential target genes in head and neck cancer. The author used the dataset GSE58911 from GEO. An interactive web tool called GEO2R, which is by default available on GEO, was used to identify the DEGs. The cutoff criteria were an adjusted P value $< 0.05$ and a log fold-change (FC) $\geq 1$ or $\leq -1$. KEGG and GO enrichment analysis was performed using the DEGs. Additionally, a PPI network was constructed, and hub genes were identified using the DEGs.
\cite{ye2021identification} Bioinformatics analysis to identify relevant HNSCC (head and neck squamous cell carcinoma) genes from public databases. In this research work, DEGs were deemed significant if their logFC $\geq 1$ or $\leq -1$ and an adjusted P value $< 0.05$. After retrieving the DEGs, several analyses were conducted, including GO, KEGG, PPI, DEG survival analyses, verification of key genes via Oncomine, specimens, and real-time PCR.
\cite{jin2019identification} Using integrated Bioinformatics techniques, the identification and analysis of genes linked to head and neck squamous cell carcinoma. GSE13601, GSE31056, and GSE30784 datasets from GEO were downloaded and analyzed using R language. To identify the DEGs from 3 different datasets, p ($<$ .01) and $|log(FC)|$ ($> 1$) were chosen as the cutoff. Common genes of the 3 different datasets were also identified by using the Venn Diagram package in R language. Further analysis (Analysis of KEGG pathways and gene ontologies, Top modules and hub genes in a PPI network identification, Validation of hub gene relative mRNA expression levels, Examining the hub genes' protein levels in the human protein atlas database, Hub gene survival analysis using the TCGA database, RNA extraction and real-time quantitative PCR, and analysis of statistics) was done based on the common DEGs.
\cite{li2020identification} By Using Integrated Bioinformatics Analysis, Hub Genes Associated With the Development of Head and Neck Squamous Cell Carcinoma Have Been Found. From TCGA and GEO, the gene expression profiles for HNSCC were retrieved. Using WGCNA, Key Co-expression Modules were identified, and DEGs were defined as genes with the cut-off criteria of $|logFC| \geq 1.0$ and adj.~$P < 0.05$. Functional Analysis of Interest Genes, PPI construction and hub gene screening, Validation of Hub Gene Expression Patterns and Prognostic Values, and Validation of Survival-Related Hub Gene Protein Expressions by the HPA Database were also conducted in this study.
\cite{shen2019identification} Using TCGA and GEO Datasets, a study was conducted to identify potential biomarkers and analyze survival data for head and neck squamous cell carcinoma. Using R, the GSE6631 dataset for head and neck squamous cell carcinoma was analyzed. Here, adj. p-val($<$ 0.05) was applied to differential gene screening in order to control the number of false positive results. The heat and volcano maps were also constructed for the corresponding DEGs. Enrichment analysis, PPI analysis, Hub genes survival analysis, Key genes verifications, analysis of Immunohistochemical, and Finding Potential Small Molecules were also identified in this paper. 

\cite{tu2021exploration} Using bioinformatics to investigate lung adenocarcinoma prognostic biomarkers. GSE31210, GSE32665, GSE32863, GSE43458, and GSE72094 datasets for lung adenocarcinoma were used in this paper. $|\log_2FC| \geq 1.5$ and $p < 0.05$ were the cut-off criteria to retrieve the DEGs from these datasets. Enrichment analysis, Finding and Verifying the Prognostic Gene Signature, Interactive Analysis of Gene Expression, Analysis of the Prognostic Model's Independence, and Nomogram construction were all conducted in this paper.
\cite{ni2019identification} The use of bioinformatics to identify important biomarkers in patients with lung adenocarcinoma. The GSE10072 dataset from the GEO database was used in this paper. The adjusted P value $< 0.05$ and $|\log_2FC| \geq 1$ were the cut-off criterion to retrieve DEGs. All of these steps such as the Analysis of KEGG pathways and gene ontologies, the Top 5 upregulated and top 5 downregulated comparison, The top 5 downregulated and top 5 upregulated stages of overall survival (OS), Analysis of the PPI network and modules were done here.
\cite{guo2019bioinformatics} Microarray data analysis using bioinformatics to find potential lung adenocarcinoma biomarkers. The GEO database was used to download the datasets (GSE118370, GSE32863, GSE85841, and GSE43458) for lung adenocarcinoma. DEGs were defined with $|\log2 Fold Change| \geq 1$ and  FDR $< 0.05$ . Analysis of GO and KEGG enrichment, Analyzing modules and building a PPI network, and Analyses of hub genes were examined here.
\cite{zhang2018elevated} Using bioinformatics analysis, elevated mRNA levels of the genes AURKA, CDC20, and TPX2 are linked to a poor prognosis for lung adenocarcinoma caused by smoking. GSE31210, GSE32863, GSE40791, GSE43458 and GSE75037 datasets from the GEO database were analyzed here. Using the cut-off criteria of P ($<$ 0.05) and absolute fold change ($>$ 1.5), the DEGs were retrieved. The functional enrichment analysis was done for 58 DEGs. After that, the Validation of data and statistical analysis steps were performed in this research work.

\section{Methodology}
\label{sec:Method}
In this section, we present the methodology of our experiments. We have introduced a process of designing gene superintendent interaction networks, including PPI networks, Interaction between TFs and genes, Network regulating gene-miRNA Interactions, and Network of the Gene-Diseases for 
Hypopharyngeal Cancer and EGFR-mutated Lung Adenocarcinoma, and also suggested common drug compounds for these two associated diseases.

The steps in the proposed methodology are described below.

\subsection{Dataset Selection}

NCBI (National Center for Biotechnology Information) is an online platform from which we can collect many forms of biological data in a variety of formats; these data are also accessible in a variety of computer-readable formats. Datasets used in this research were gathered from the NCBI platform's GEO (Gene Expression Omnibus) database. The GEO database for high throughput gene expression analysis can be accessed through the National Center for Biotechnology Information platform \cite{edgar2002gene}. RELA is dependent on CD271 expression and stem-like features in Hypopharyngeal cancer, according to the dataset (GSE212398). 
The dataset (GSE198672) contains EGFR-mutated lung adenocarcinomas that develop from pre-existing tumor cells and persist in a specialized stromal milieu as drug-tolerant persisters after erlotinib treatment. The RNA Sequence from GSE198672 and GSE212398 was extracted using the GPL10558 (Illumina HumanHT-12 V4.0 expression bead chip) and GPL20844 (Agilent-072363 SurePrint G3 Human GE v3 8x60K Microarray 039494 [Feature Number Version]) platforms, respectively. The GSE212398 dataset is a subseries of the GSE212399 dataset. For our investigation, we chose the GSE212398 dataset because this dataset contains 8 samples including 4 samples for Control and 4 samples for KO. The GSE198672 dataset also has 8 samples. 

\subsection{Differential Expression analysis}

Finding Differentially Expressed Genes(DEGs) from these microarray datasets is the initial step for this particular research. The \textit{GEOquery} \cite{davis2007geoquery} R package was used for retrieving gene expression datasets for both diseases from the NCBI GEO \cite{clough2016gene} database. Next, the \textit{limma} \cite{smyth2005limma} R package with empirical Bayes statistics was used for differential expression (DE) analysis. The DE output was formatted as a comma-separated values (CSV) file containing information, including Gene Symbol, logFC, p.value, and adjusted p.Val for the corresponding disease dataset and collected. Both datasets' false discovery rates were controlled using the Benjamini-Hochberg \cite{benjamini1995controlling} method. An adjusted P-value ($<$ 0.05) and log2-fold change (absolute) > 1 are the cutoff criteria for obtaining DEGs for both datasets. Using the Venny tool \cite{oliveros2007venny}, the shared DEGs between these two diseases were visualized.

\subsection{Enrichment analysis of shared DEGs}

Enrichment of gene sets is the study of gene sets with connected chromosomal locations, molecular activities, and biological functions \cite{subramanian2005gene}. Gene Ontology (GO), which is divided into the three categories of biological process, molecular function, and cellular component, is used for gene product annotation \cite{doms2005gopubmed}. Understanding the molecular activity, cellular function, and the location in a cell where the genes perform their functions serves as the main foundation for choosing GO keywords \cite{taz2021network}. The Kyoto Encyclopedia of Genes and Genomes (KEGG) pathway, which has a significant advantage over gene annotation, is frequently used to study metabolic pathways \cite{kanehisa2000kegg}. For extensive route analysis, databases from Reactome \cite{fabregat2018reactome}, BioCarta \cite{nishimura2001biocarta}, and WikiPathways \cite{slenter2018wikipathways} were used in addition to the KEGG pathway.

\subsection{TF-miRNA coregulatory network}
To determine which transcription factors (TFs) bind with shared DEGs in the regulatory regions, target gene relationships between TFs and TFs were examined \cite{al2020detection}. MiRNAs that attempt to bind on a gene transcript to negatively affect protein expression have been identified using miRNAs target gene interactions \cite{hsu2011mirtarbase}. The RegNetwork repository \cite{liu2015regnetwork} provided interactions for TF-miRNA coregulatory interactions, which make it easier to identify regulatory TFs and miRNAs that regulate DEGs of interest during the transcriptional and post-transcriptional phases \cite{taz2021network}.  Utilizing the NetworkAnalyst platform, we constructed the TF-miRNA coregulatory network \cite{zhou2019networkanalyst}. Researchers can browse complex datasets with the help of NetworkAnalyst to find biological traits and functionalities that can be used to generate useful biological hypotheses \cite{xia2015networkanalyst}. The minimum Network option was selected among the different available formats to construct the TF-miRNA coregulatory network.

\subsection{PPI Network}
PPI activity is thought to be the main area of interest in cellular biology research and is necessary for system biology \cite{ewing2007large}. With the aid of cutting-edge research on PPI networks, the number of complex biological processes is identified \cite{pagel2005mips} \cite{chowdhury2020network}. Proteins operate inside of cells through interactions with other proteins, and information produced by a PPI network aids in our understanding of how proteins function \cite{ben2005kernel}. Given to the STRING \cite{mering2003string} database  (https://string-db.org/),  shared DEGs of Hypopharyngeal cancer and EGFR-mutated lung adenocarcinoma are used to create a PPI network and discover the genes that are directly associated among the common genes. Some basic settings are set on the STRING to get our desired result such as setting the network type as full STRING network, selecting the meaning of network edges by evidence, also selecting active interaction sources by Textmining, Experiments, Databases, Co‑expression, Neighborhood, Gene Fusion, and Co‑occurrence. Interactive svg (network is a scalable vector graphic [SVG]; interactive ) is selected for network display mode in Advanced Settings. The information provided by STRING is based on expected and experimental interactions, and the interactions generated by the web tool are characterized by 3D structures, supplementary data, and evidence scores \cite{szklarczyk2010string}. After constructing this PPI network from STRING, this STRING PPI network was further reconstructed in Cytoscape to identify only the interconnected Genes and remove the disconnected genes among those 32 shared genes. With the help of the web-based NetworkAnalyst \cite{zhou2019networkanalyst} software  (https://www.networkanalyst.ca/),  identified directly interconnected genes (from Cytoscape  PPIs network) were entered into InnateDB \cite{breuer2013innatedb} to design additional PPIs for interconnected genes. Here Auto Layout was selected in the Layout option to build this network.
 
\subsection{Retrieving Hub Genes}
Hub nodes are referred to be the highly connected nodes in a large-scale PPI network \cite{hsing2008use}. The cytoHubba plugin for the Cytoscape program is used to locate hub nodes. The user-friendly cytoHubba interface makes it the most popular hub identification plugin for Cytoscape, and it comes with 11 topological analysis methods \cite{chin2014cytohubba}.  Among the 11 topological methods on CytoHubba, The Degree method and The maximum clique centrality (MCC) were chosen to identify the hub genes. In the degree topology method, the degree is counted according to the number of interactions among the genes. Higher interacted genes from the given input genes are easily identified. The gene has the highest number of degree scores ranking as top among all genes. The most important candidate genes among the shared DEGs, which may be crucial in physiological regulatory functions, were found using the maximal clique centrality (MCC), which demonstrated better accuracy in predicting critical proteins in the PPI network. 

The Maximal Clique Centrality (MCC) technique was found to be the most efficient way to locate hub nodes in a PPI network \cite{shi2021identification}. Also the authors of article \cite{wang2021identification} mentioned that the most efficient technique for identifying hub nodes was thought to be the Maximal Clique Centrality (MCC) algorithm. So, these two methods (The Degree Topology Method, One of the most popular topological method \cite{vallabhajosyula2009identifying} and The maximum clique centrality (MCC), the most efficient method among the available 11 methods \cite{chin2014cytohubba}) that were chosen to identify hub genes out of the 11 available. 

\subsection{Functional association Network}
A Bioinformatics program called GeneMANIA displays functional association information, genetic relationships, pathways, and co-expression for a given set of input data \cite{habib2017drug}.  Gene sets' functions can be predicted with the aid of GeneMANia \cite{zuberi2013genemania}. The percentage of Co-expression, Physical interactions, Predicted, Pathway, Co-localization, Shared protein domains and Genetic interactions for the given input genes are easily identified through the Functional association Network. Physical interactions between two or more proteins can produce binary interactions and complex proteins \cite{frishman2009modern}. Genes are associated in gene co-expression networks, which are transcription factor-transcription factor association networks that are typically presented as undirected graphs \cite{tieri2019network}. Unlike most co-expression networks, which are undirected graphs, this network showed a close relationship between two genes \cite{ruan2010general}. In studies on protein-protein interactions, two genes are connected if they are found to interact. These ligand-based protein networks, which foresee the ability of nearby proteins to bind connected substances indirectly, may be used to enhance genetically orientated gene networks, which foretell the significance of a procedure or a disease \cite{hasan2020design}. Using a data analysis technique called gene co-expression analysis, it is possible to find groupings of genes that have comparable expression patterns under various conditions \cite{roy2014reconstruction}. The link between the genes' functions is referred to as genetic interaction \cite{basar2023identification}. The top 10 hub genes were used to demonstrate a functional network from GeneMania \cite{franz2018genemania}. 

\subsection{TF-gene interactions}

By analyzing the TF-gene interaction using the discovered 10 hub genes, one may determine the impact of TF on functional pathways and gene expression levels \cite{ye2019bioinformatic}. Users can do a meta-analysis and analyze gene expression for numerous species with the use of NetworkAnalyst \cite{zhou2019networkanalyst}.  The control of gene transcription as well as the establishment of cellular identity and activity are assumed to depend on transcriptional factors (TFs), the TF (transferrin) gene-producing proteins \cite{ye2017integrative}. The TF-gene interaction investigates how TF affects functional pathways and levels of gene expression \cite{basar2023system}. Finding the important TF-gene interactions is crucial for comprehending the roles of pleiotropic global regulators \cite{tran2010trimming}. Through direct or indirect interactions with other TFs, specific TFs help regulate the expression of a variety of target genes \cite{ye2017integrative}. To control life activities, several transcription factors interact \cite{kazemian2013widespread}. The 10 hub genes are utilized to evaluate the impact of TF on the functional pathways and expression levels of the genes through TF-gene interaction analysis. To find TF-gene interactions with well-known genes, researchers use the NetworkAnalyst platform \cite{zhou2019networkanalyst}. NetworkAnalyst includes activities that are typical of network topologies and can be used to analyze biological modules \cite{xia2014networkanalyst}.  The NetworkAnalyst platform's ChEA \cite{lachmann2010chea} database provided inspiration for the network built for the TF-gene interaction network \cite{taz2021network}.

\subsection{Gene-miRNA interactions}

 By analyzing the Gene-miRNA interaction using the discovered 10 hub genes, one may determine the impact of TF on functional pathways and gene expression levels By base-pairing with their target mRNAs, microRNAs, a class of brief, non-coding RNA molecules with a length of 21–25 nucleotides, regulate the expression of genes, primarily by silencing or down-regulating the target genes \cite{leon2017visualization}.   Natural single-stranded tiny RNA molecules known as microRNAs control the expression of genes by attaching to certain mRNAs and either starting the translation of the target mRNA or starting the destruction of the target mRNA \cite{kuhn2008experimental}. Small non-coding RNAs known as microRNAs (miRNAs) were discovered to promote mRNA degradation or prevent post-transcriptional translation \cite{kuhn2008experimental}. Evidence is mounting that miRNAs have a role in carcinogenesis and cancer metastasis \cite{lynam2009roles}.  
 More and more varieties and uses for small non-coding RNAs are being discovered. This implies that there may be regulatory mechanisms that are far more complex than those now employed in the analysis and creation of gene regulatory networks. Finding new therapeutic targets can benefit from the analysis of inter-pathway regulatory factors. Since non-coding miRNAs are important for activating pathways, their activity is crucial in this regulatory environment. In the control of transcriptome processes, microRNAs are crucial \cite{shah2010profiling}. For many biological processes in both plants and animals, posttranscriptional mediators of gene expression such as microRNAs are crucial \cite{carthew2006gene}. To fully comprehend a miRNA's biological function, it is crucial to pinpoint the genes that it regulates \cite{basar2023identification}. MiRNAs can be retrieved by using the TarBase database. TarBase is a comprehensive repository of animal microRNA targets supported by experimental data. The database is also functionally connected to several other helpful resources, including Gene Ontology (GO) and the UCSC Genome Browser. TarBase provides a rich data set from which to evaluate characteristics of miRNA targeting that will be helpful for the upcoming generation of target prediction tools. TarBase reveals substantially more empirically supported targets than even recent evaluations indicate \cite{karagkouni2018diana}. The network of gene-miRNA interactions is created using the web-based tool TarBase under NetworkAnalyst for those 10 hub genes (JUN, ERBB2, HLA-DMB, HBEGF, HLA-B, HLA-DRA, DUSP5, ARHGDIB, MUC4, CLEC2D). 

\subsection{Gene-disease interactions}
 Gene-disease interactions network focuses mostly on the most recent understanding of human genetic illnesses, including complex, mendelian, and ecological diseases \cite{pinero2016disgenet}.   Gene-disease interactions network helps to identify those diseases that can occur due to the input genes. This network helps us to identify the risk factors that should be cured by therapeutic molecules. Gene-disease interactions network focuses primarily on the most recent knowledge of complex and ecological diseases, as well as other human genetic ailments \cite{bauer2011gene}.  DisGeNET is a sizable database of gene-disease interactions that combines information from several sources and covers a range of biological traits linked to diseases \cite{pinero2016disgenet}. The hub genes were linked to related diseases and their chronic states by the network analysis of gene-disease correlations. DisGeNET \cite{pinero2016disgenet} is a large database of gene-disease interactions that incorporates links from several sources and covers a variety of biological features associated with illnesses \cite{al2020detection}. The investigation of gene-disease correlations using NetworkAnalyst identified associated diseases and their chronic conditions with the hub genes (JUN, ERBB2, HLA-DMB, HBEGF, HLA-B, HLA-DRA, DUSP5, ARHGDIB, MUC4, CLEC2D) \cite{xia2015networkanalyst}.

\subsection{Therapeutic molecule suggestion for selected diseases}
 Therapeutic molecule suggestion is the pointer step of this research. DSigDB is used for drug suggestion. Users can access the DSigDB database via the Enrichr platform (https://amp.pharm.mssm.edu/Enrichr/)  \cite{taz2021network}. Enrichr is mostly used as an enrichment analysis tool, which offers substantial graphical data on the combined functions of the input genes \cite{chen2013enrichr}. There are 19,531 genes, 22,527 gene sets, and 17,389 unique chemicals in DSigDB \cite{yoo2015dsigdb}.  A new gene set resource called Drug Signatures Database (DSigDB) connects medicines and compounds with their target genes.   To forecast drugs, DSigDB largely employs gene expression-based datasets, and each group of genes is seen as being targeted when taking a molecule into account \cite{yoo2015dsigdb}. 

\section{Results}
\label{results}

\subsection{Differential Expression analysis identifies common DEGs between Hypopharyngeal cancer and EGFR-mutated lung adenocarcinoma}
\begin{figure}[!htbp]
    \centering
    \includegraphics[scale=.2]{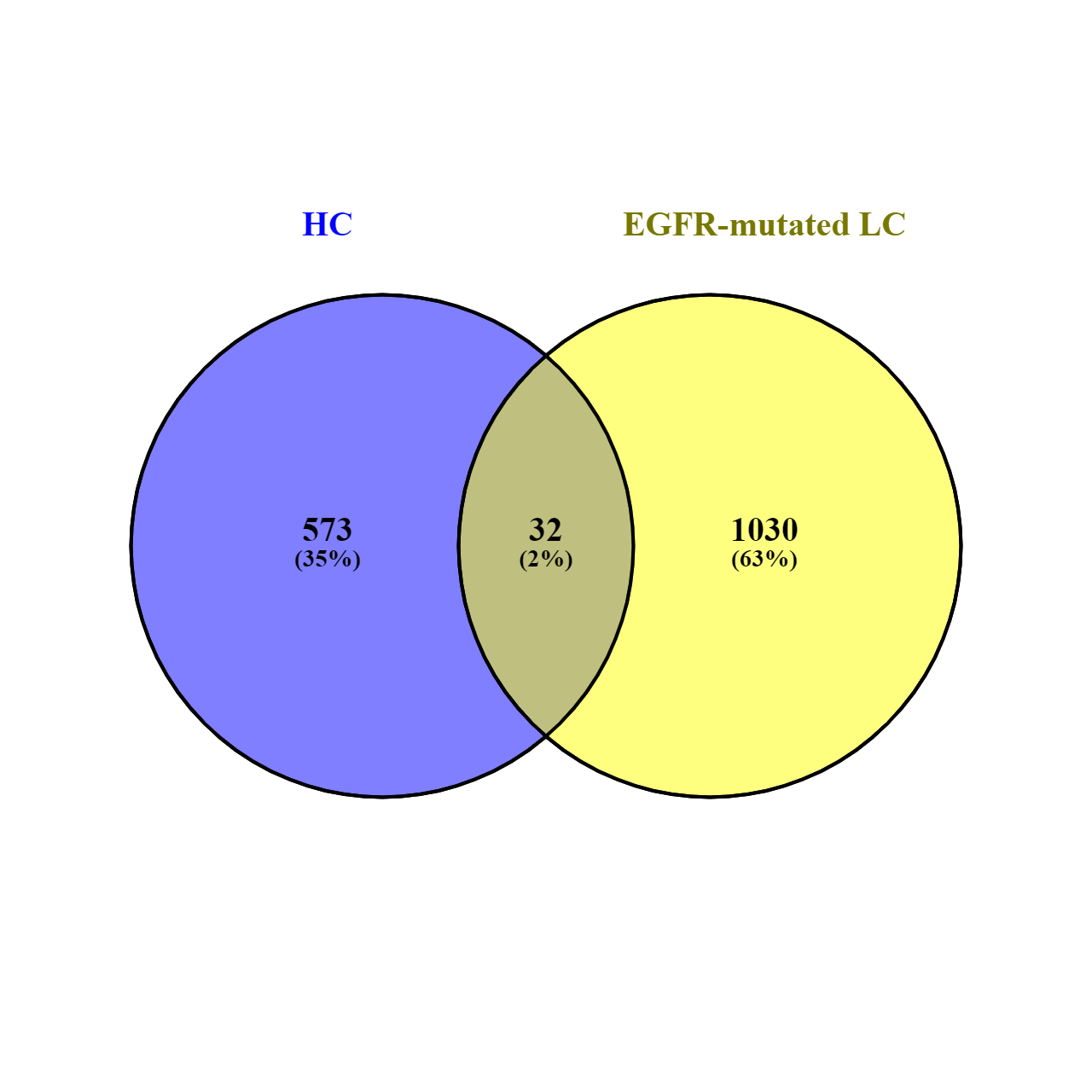}
    \caption{Venn Diagram of shared DEGs. 32 common genes were found between HC and EGFR-mutated LC. Common DEGs were 2\% among 1667 DEGs.}
    \label{fig:Venn}
\end{figure}

We found 605 identical DEGs for Hypopharyngeal cancer (GSE212398) and 1062 identical DEGs for EGFR-mutated lung adenocarcinoma (GSE198672) by using the R programming language. Among those identical DEGS, 32 common genes were identified between Hypopharyngeal cancer and EGFR-mutated lung adenocarcinoma through the Venny tool. The ven Diagram of Shared DEGs between the two diseases is shown in Figure\ref{fig:Venn}. 

\subsection{Enrichment of functional pathways and Gene ontology terms}
\begin{figure*}[!htbp]
	\centering
     \subfloat[biological process]{\includegraphics[scale=.35]{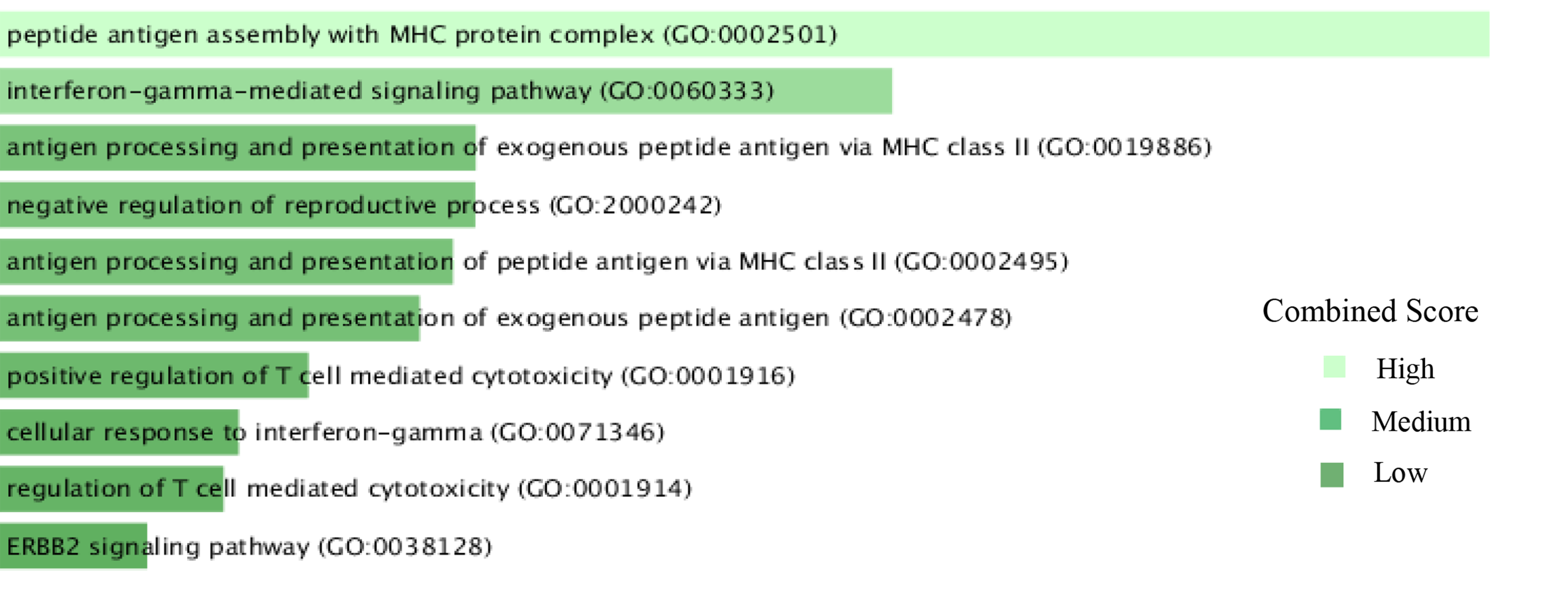}}\hspace{0.3cm}
	\subfloat[molecular function]{\includegraphics[scale=.35]{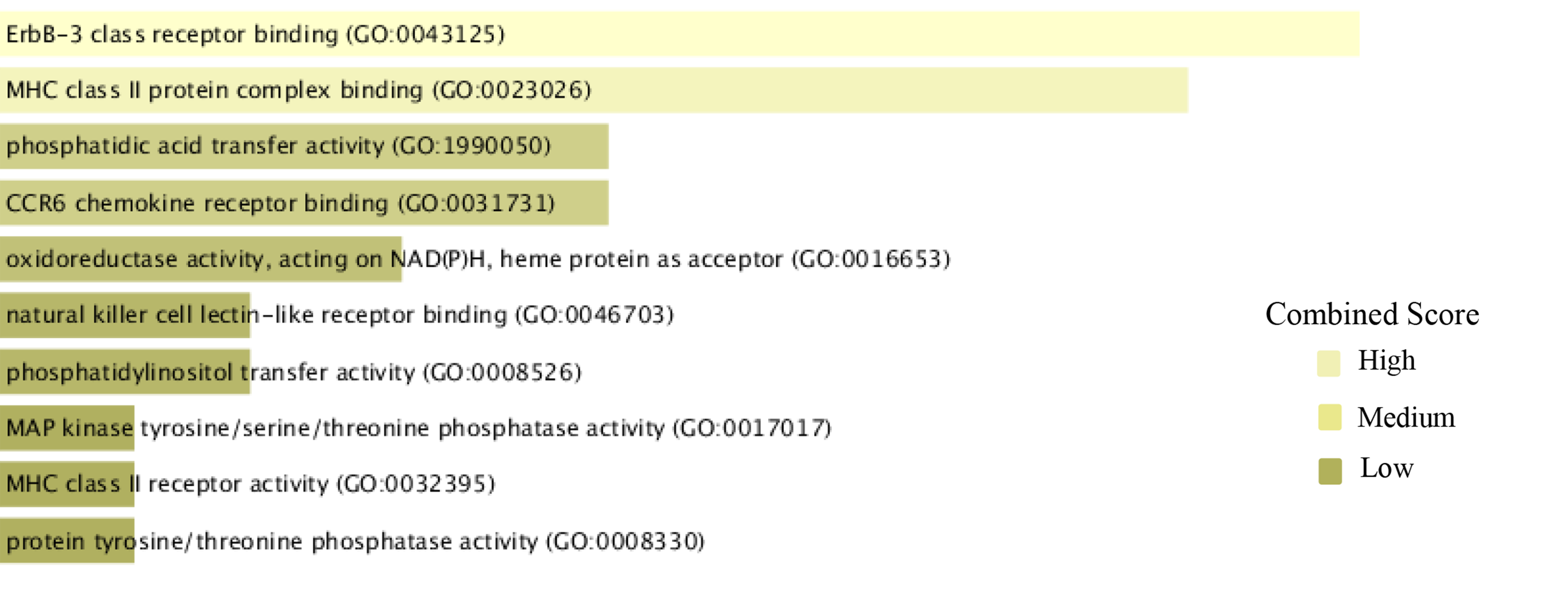}}
 \hspace{0.3cm}
	\subfloat[cellular component]{\includegraphics[scale=.35]{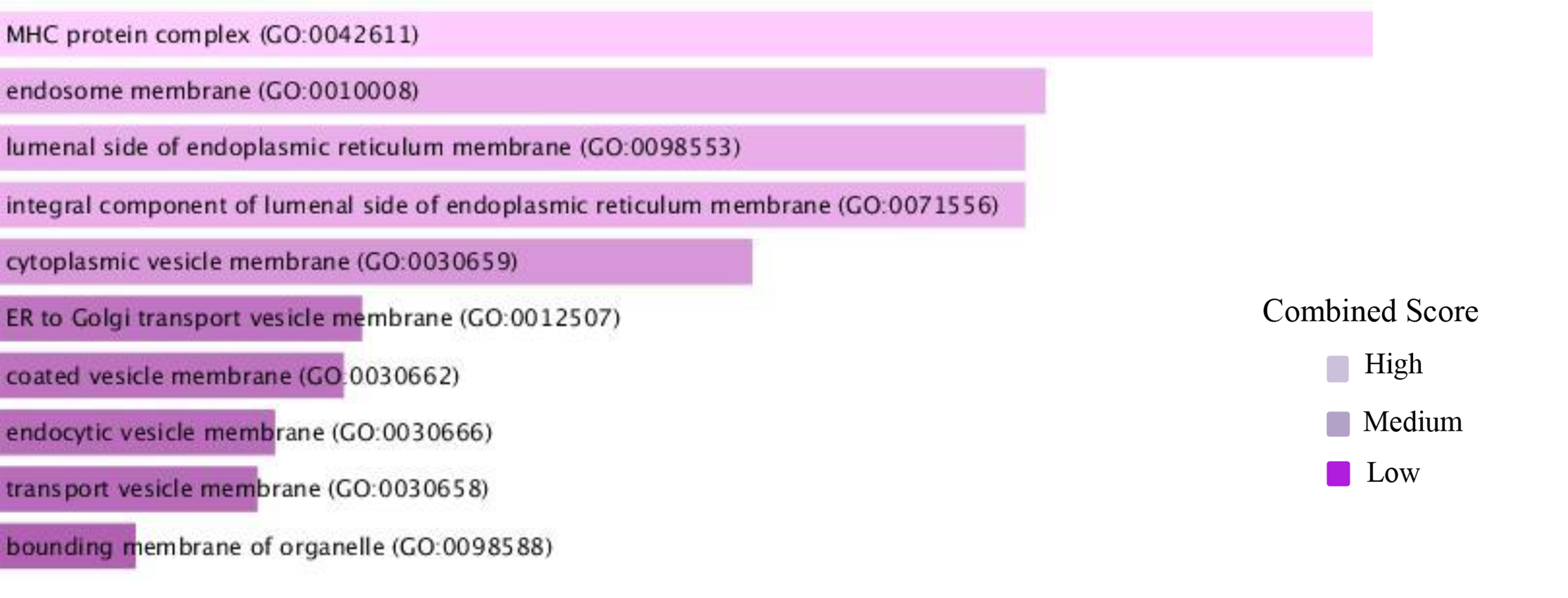}}
 \hspace{0.1cm}
	\caption{Top 10 GO terms concomitant to biological process, molecular function, and cellular component pinpointing entrenched on the combined score (The log of the p-value from the Fisher exact test and multiplying that by the z-score of the deviation from the expected rank.}
	\label{fig:CNN31}
\end{figure*}

\begin{figure*}[!htbp]
	\centering
     \subfloat[Reactome]{\includegraphics[scale=.28]{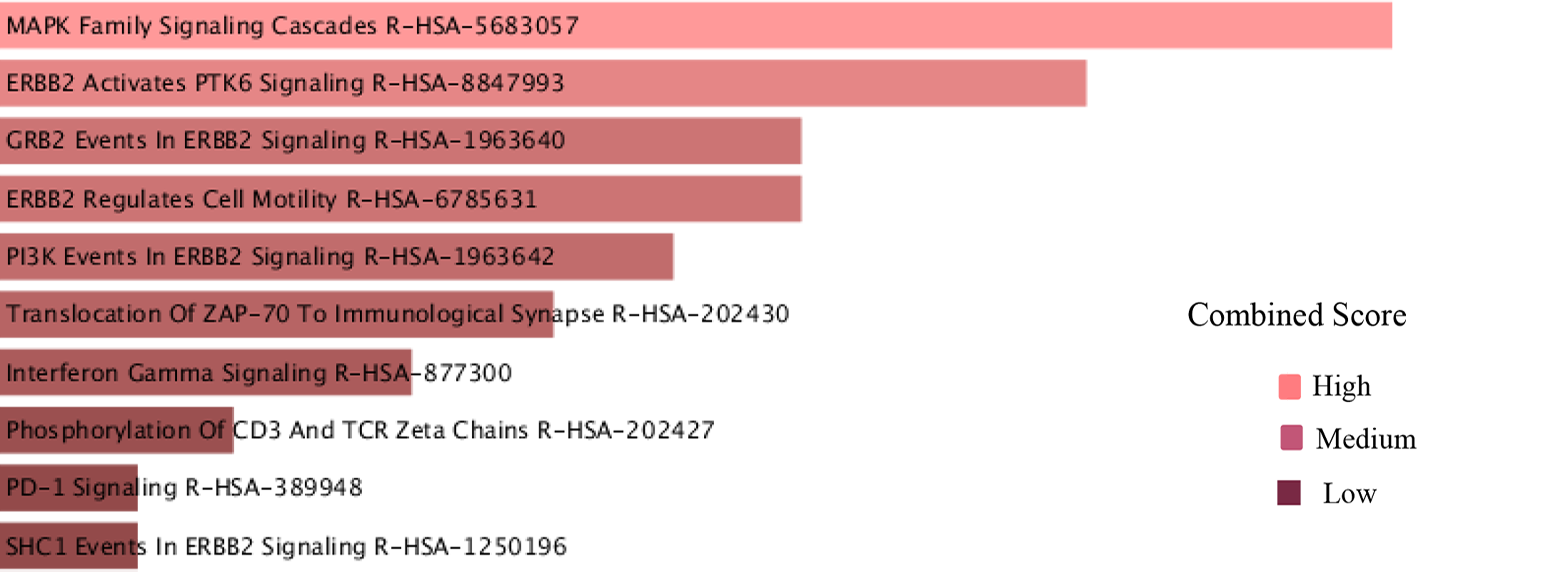}}
     \hspace{0.3cm}
     
	\subfloat[KEGG]{\includegraphics[scale=.28]{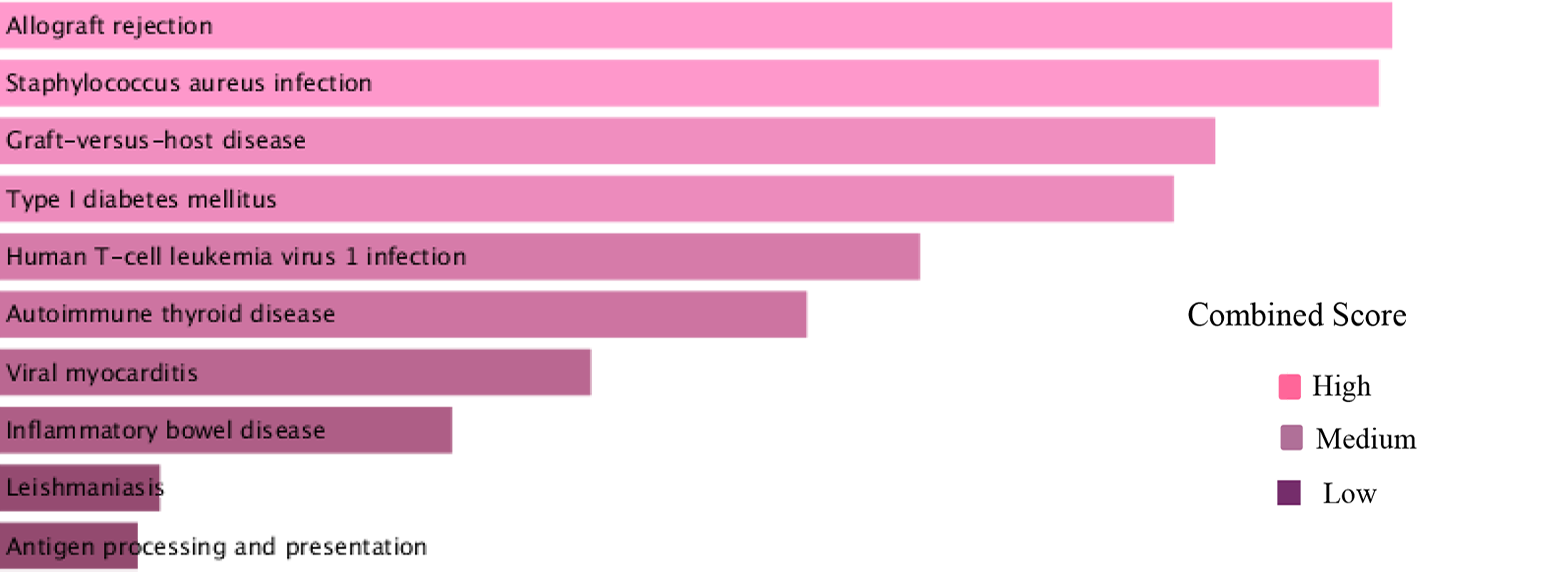}}
 \hspace{0.3cm}
    
	\subfloat[WikiPathway]{\includegraphics[scale=.28]{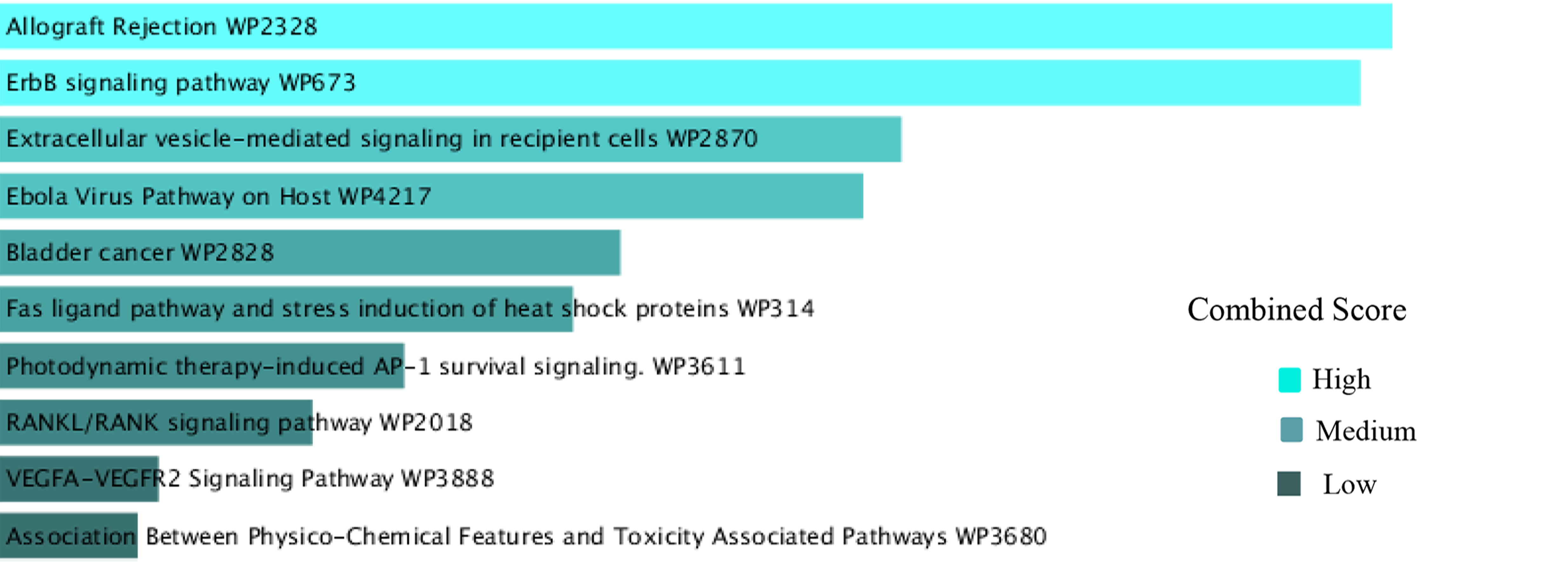}}
 \hspace{0.3cm}
    
	\subfloat[BioCarta]{\includegraphics[scale=.30]{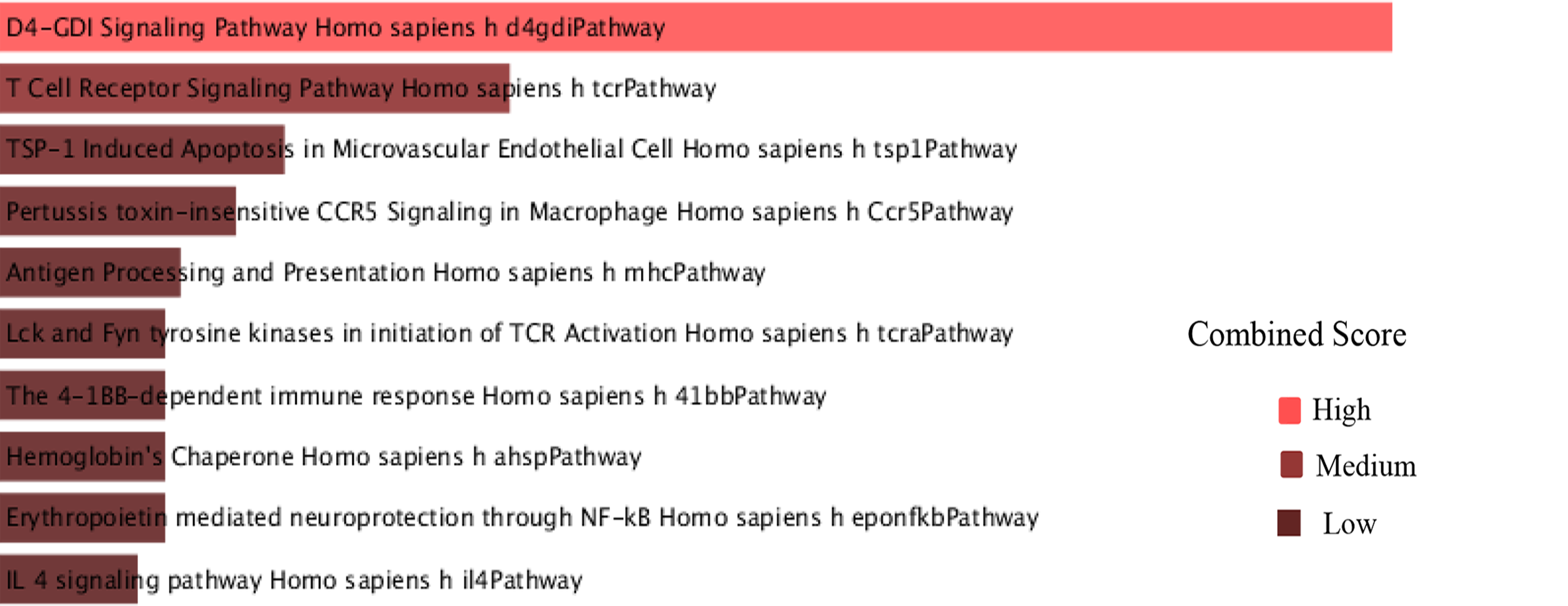}}
 \hspace{0.1cm}
    \caption{Top 10 pathways from (a) Reactome, (b) KEGG, (c) WikiPathways, and (d) BioCarta pinpointing entrenched on the combined score (The log of the p-value from the Fisher exact test and multiplying that by the z-score of the deviation from the expected rank).}
	\label{fig:CNN32}
\end{figure*}

The analysis of gene set enrichment was performed using the online tool Enrichr \cite{taz2021network}. Many databases, including The GO \cite{gene2004gene}, Reactome \cite {fabregat2018reactome}, KEGG \cite {kanehisa2007kegg}, WikiPathways \cite{ martens2021wikipathways}, and BioCarta \cite {nishimura2001biocarta}, were used \cite{taz2021identification} to find GO keywords and cell-informing pathways. The GO database was used to find the biological process, molecular function, and cellular components. Analysis of biological process, molecular function, and cellular component data revealed notable involvement in Peptide antigen assembly with MHC protein complex, ErbB-3 class receptor binding, and MHC protein complex in shared DEGs respectively. MAPK Family Signaling Cascades, Allograft rejection, Allograft Rejection, and D4-GDI Signaling Pathway were highly enriched among all identified when Reactome, KEGG, WikiPathways, and BioCarta databases were used, respectively. Table \ref{table: GO terms} shows the top 5 Biological terms, Cellular terms, and Molecular terms and table \ref{table: pathway} shows the top 5 pathways from Reactome, KEGG, WikiPathways, and BioCarta with correspondent P-value and genes. Top 10 GO terms concomitant to biological process, molecular function, and cellular component pinpointing entrenched on combined score in this Figure \ref{fig:CNN31} and also based on a combined score, the top 10 pathways from Reactome, KEGG, WikiPathways, and BioCarta are mentioned in this Figure \ref{fig:CNN32}.  To get the combination score, multiply the z score, which represents the deviation from the predicted rank, by the log of the p-value from the Fisher exact test. The “Combined scores” for Figures 3 and 4 are automatically calculated in the Enrichr platform. In the biological process, peptide antigen assembly with MHC protein complex (GO:0002501) indicates “Peptide attachment to an MHC protein complex's antigen-binding groove”. The Interferon-gamma-mediated signaling pathway (GO:0060333) means the cascade of molecular signals that begins when interferon-gamma binds to its receptor on a target cell's surface and ends with the control of a cell's transcription, among other downstream cellular processes. The only type II interferon so far discovered is interferon-gamma. The antigen processing and presentation of exogenous peptide antigen via MHC class II (GO:0019886) is the process by which an MHC class II protein complex collaborates with an antigen-presenting cell to express a peptide antigen of external origin on the cell surface. Typically, but not always, a complete protein is used to digest the peptide antigen. The negative regulation of reproductive processes (GO:2000242) indicates any procedure that slows down, prevents, or lessens the number of times, how often, or how much the reproductive process occurs. The antigen processing and presentation of peptide antigen via MHC class II (GO:0002495) is the process by which an MHC class II protein complex collaborates with an antigen-presenting cell to express a peptide antigen on the cell surface. Usually, but not always, the protein in its whole serves as the source of the peptide antigen.
In cellular components, the MHC protein complex (GO:0042611) is An MHC class II beta chain or an invariant beta2-microglobulin chain, along with or without a bound peptide, lipid, or polysaccharide antigen, makes up a transmembrane protein complex. The endosome membrane (GO:0010008) indicates a lipid bilayer that envelops an endosome. The lumenal side of the endoplasmic reticulum membrane (GO:0098553) indicates the leaflet-shaped side of the plasma membrane that is facing the lumen. The integral component of the lumenal side of the endoplasmic reticulum membrane (GO:0071556) indicates a portion of the endoplasmic reticulum membrane made up of gene products that can only pass through the membrane's lumenal side. The cytoplasmic vesicle membrane (GO:0030659) is a cytoplasmic vesicle's protective lipid bilayer. 
In molecular function, ErbB-3 class receptor binding (GO:0043125) indicates ErbB-3/HER3 protein-tyrosine kinase receptor binding. The MHC class II protein complex binding (GO:0023026) is the main histocompatibility complex of class II. The phosphatidic acid transfer activity (GO:1990050) means phosphatidic acid is taken out of a membrane or a monolayer lipid particle, transported through the aqueous phase while being sheltered in a hydrophobic pocket, and then brought to a membrane or lipid particle that will accept it. Phosphatidic acid is a type of glycophospholipid that typically has a phosphate group attached to carbon-3, an unsaturated fatty acid attached to carbon-2, and a saturated fatty acid attached to carbon-1. The CCR6 chemokine receptor binding (GO:0031731) is chemokine CCR6 receptor binding. The oxidoreductase activity, acting on NAD(P)H, heme protein as acceptor (GO:0016653) indicates an oxidation-reduction (redox) reaction that uses NADH or NADPH as a hydrogen or electron donor to reduce a heme protein is catalyzed.

\begin{table}[!htbp]
    \begin{center}
    \caption{Biological terms, Cellular terms, and Molecular terms with correspondent P-value and genes}
    \label{table: GO terms}
        \resizebox{\textwidth}{!}{

        \begin{tabular}{ |p{2.2cm}|p{9cm}|l|p{5cm}| }

        \hline
           Type & Term & P-value & Genes\\
             \hline
             GO Biological Process & peptide antigen assembly with MHC protein complex (GO:0002501) & 3.71E-05 & HLA-DMB;HLA-DRA \\
             & interferon-gamma-mediated signaling pathway (GO:0060333) & 1.74E-04 & HLA-DRB4;HLA-B;HLA-DRA \\
             & antigen processing and presentation of exogenous peptide antigen via MHC class II (GO:0019886) & 5.10E-04 & HLA-DMB; HLA-DRB4; HLA-DRA \\
             & negative regulation of reproductive process (GO:2000242) & 5.11E-04 & ARHGDIB;NKX3-1\\
             & antigen processing and presentation of peptide antigen via MHC class II (GO:0002495) & 5.41E-04 & HLA-DMB; HLA-DRB4; HLA-DRA \\
             \hline
             GO Cellular Component & MHC protein complex (GO:0042611) & 4.16E-06 & HLA-DMB;HLA-B;HLA-DRA \\
             & endosome membrane (GO:0010008) & 1.12E-05 & HLA-DMB;HLA-DRB4;ERBB2;HLA-B;HLA-DRA;RHOD \\
             & lumenal side of endoplasmic reticulum membrane (GO:0098553) & 1.19E-05 & HLA-DRB4;HLA-B;HLA-DRA \\
             & integral component of lumenal side of endoplasmic reticulum membrane (GO:0071556) & 1.19E-05 & HLA-DRB4;HLA-B;HLA-DRA\\
             & cytoplasmic vesicle membrane (GO:0030659) & 2.70E-05 & HLA-DRB4;ERBB2;HLA-B;HLA-DRA;RHOD;HBEGF \\
             \hline
             GO Molecular Function & ErbB-3 class receptor binding (GO:0043125) & 0.007975 & ERBB2 \\
             & MHC class II protein complex binding (GO:0023026) & 3.32E-04 & HLA-DMB;HLA-DRA \\
             & phosphatidic acid transfer activity (GO:1990050) & 0.011148 & PITPNC1 \\
             & CCR6 chemokine receptor binding (GO:0031731) & 0.011148 & DEFB1 \\
             & oxidoreductase activity, acting on NAD(P)H, heme protein as acceptor (GO:0016653) & 0.012731 & CYB5R2 \\
             \hline
             \end{tabular}
             }
    \end{center}
\end{table}

\begin{table}[]
\caption{5 pathways from Reactome, KEGG, WikiPathways, and BioCarta and with correspondent P-value and genes}
\label{table: pathway}
\resizebox{\textwidth}{!}{%

\begin{tabular}{|l|p{7cm}|l|p{6cm}|}
\hline
Database              & Pathway                                                       & P-value     & Genes                                \\ \hline
\multirow{5}{*}{Reactome} &
  MAPK Family Signaling Cascades R-HSA-5683057 &
  1.39E-04 &
  DUSP5;JUN;ERBB2; PEA15;HBEGF \\ \cline{2-4} 
                      & ERBB2 Activates PTK6 Signaling R-HSA-8847993                  & 1.91E-04    & ERBB2;HBEGF                          \\ \cline{2-4} 
                      & GRB2 Events In ERBB2 Signaling R-HSA-1963640                  & 2.57E-04    & ERBB2;HBEGF                          \\ \cline{2-4} 
                      & ERBB2 Regulates Cell Motility R-HSA-6785631                   & 2.57E-04    & ERBB2;HBEGF                          \\ \cline{2-4} 
                      & PI3K Events In ERBB2 Signaling R-HSA-1963642                  & 2.93E-04    & ERBB2;HBEGF                          \\ \hline
\multirow{5}{*}{}{KEGG} & Allograft rejection                                           & 3.83E-07    & HLA-DMB;HLA-DRB4;HLA-B;HLA-DRA       \\ \cline{2-4} 
                      & Staphylococcus aureus infection                               & 3.96E-07    & HLA-DMB;HLA-DRB4;KRT16;HLA-DRA;DEFB1 \\ \cline{2-4} 
                      & Graft-versus-host disease                                     & 5.79E-07    & HLA-DMB;HLA-DRB4;HLA-B;HLA-DRA       \\ \cline{2-4} 
                      & Type I diabetes mellitus                                      & 6.37E-07    & HLA-DRB4;HLA-B;HLA-DRA               \\ \cline{2-4} 
 &
  Human T-cell leukemia virus 1 infection &
  1.15E-06 &
  JUN;EGR2;HLA-DMB;HLA-DRB4;HLA-B;HLA-DRA \\ \hline
\multirow{5}{*}{WikiPathway} &
  Allograft Rejection WP2328 &
  3.85E-04 &
  HLA-DMB;HLA-B;HLA-DRA \\ \cline{2-4} 
                      & ErbB signaling pathway WP673                                  & 4.11E-04    & JUN;ERBB2;HBEGF                      \\ \cline{2-4} 
 &
  Extracellular vesicle-mediated signalling in recipient cells WP2870 &
  0.001049058 &
  TSPAN8;ERBB2 \\ \cline{2-4} 
                      & Ebola Virus Pathway on Host WP4217                            & 0.001134032 & HLA-DMB;HLA-B;HLA-DRA                \\ \cline{2-4} 
                      & Bladder cancer WP2828                                         & 0.001862402 & ERBB2;HBEGF                          \\ \hline
\multirow{5}{*}{}{BioCarta} &
  D4-GDI Signaling Pathway Homo sapiens h d4gdiPathway &
  3.71E-05 &
  JUN;ARHGDIB \\ \cline{2-4} 
 &
  T Cell Receptor Signaling Pathway Homo sapiens h tcrPathway &
  0.003493127 &
  JUN;HLA-DRA \\ \cline{2-4} 
 &
  TSP-1 Induced Apoptosis in Microvascular Endothelial Cell  Homo sapiens h tsp1Pathway &
  0.011147938 &
  JUN \\ \cline{2-4} 
 &
  Pertussis toxin-insensitive CCR5 Signaling in Macrophage Homo sapiens h Ccr5Pathway &
  0.014310901 &
  JUN \\ \cline{2-4} 
                      & Antigen Processing and Presentation Homo sapiens h mhcPathway & 0.01903698  & HLA-DRA                              \\ \hline
\end{tabular}
}
\end{table}

 \subsection{TF-miRNA coregulatory network construction}
\begin{figure}[!htbp]
    \centering
    \includegraphics[scale=0.75]{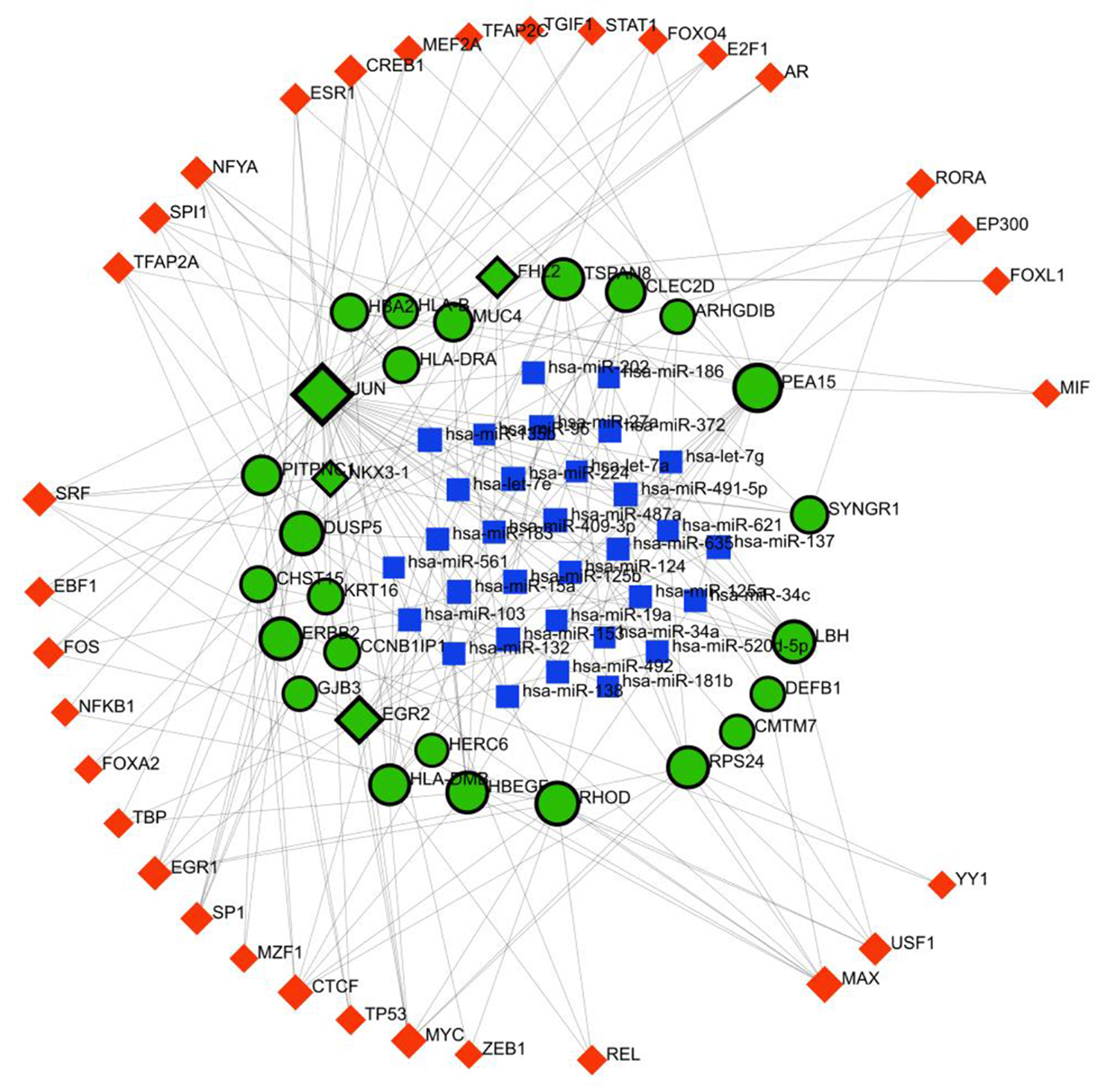}
    \caption{Visualization of TF-miRNA coregulatory network through NetworkAnalyst. Green-black highlighted Nodes indicate seeds, Red Diamond-shaped nodes for TF, and Blue Box Shaped Nodes for miRNA.}
    \label{fig: TF-miRNA}
\end{figure}

To comprehend how TF and miRNA regulate with shared DEGs, a TF-miRNA coregulatory network was developed. Common DEGs (GRAMD3, SYNGR1, CMTM7, HLA-DMB, HLA-DRB4, CLEC2D, CCNB1IP1, HLA-DRA, DEFB1, ERBB2, MUC4, LOC145837, RPS24, RHOD, HBEGF, ARHGDIB, PITPNC1, PEA15, KRT16, GJB3, JUN, FHL2, CYB5R2, HLA-B, EGR2, HERC6, DUSP5, HBA2, CHST15, NKX3-1, LBH, TSPAN8) were utilized to create the network of TF-miRNA coregulators.  32 common genes were given as “Gene Input List” in NetworkAnalyst. Then “H.sapiens (human)” and “Official Gene Symbol” were chosen for “Specify organism” and “Set ID type” attributes correspondingly.  After uploading this information, TF-miRNA coregulatory interactions were selected from Gene Regulatory Interaction.  The literature-curated regulatory interaction information was collected from the RegNetwork repository. After selecting the Minimum Network from the Network Tools option, the TF-miRNA coregulatory network was constructed. Background as White and layout as Circular Bi/Tripartite were chosen to better visualise the network. Also, Opacity, Thickness, Color, Label and size were also customized from Edge and Node options. Red color for TF, Green-Black highlighted for seeds and Blue for miRNA were chosen from The Global Node Styles.  The network shown in Figure \ref{fig: TF-miRNA}, comes with 93 nodes, 223 edges, and 28 seeds.

\subsection{PPI Network}
\begin{figure}[!htbp]
    \centering
    \includegraphics[scale=.45]{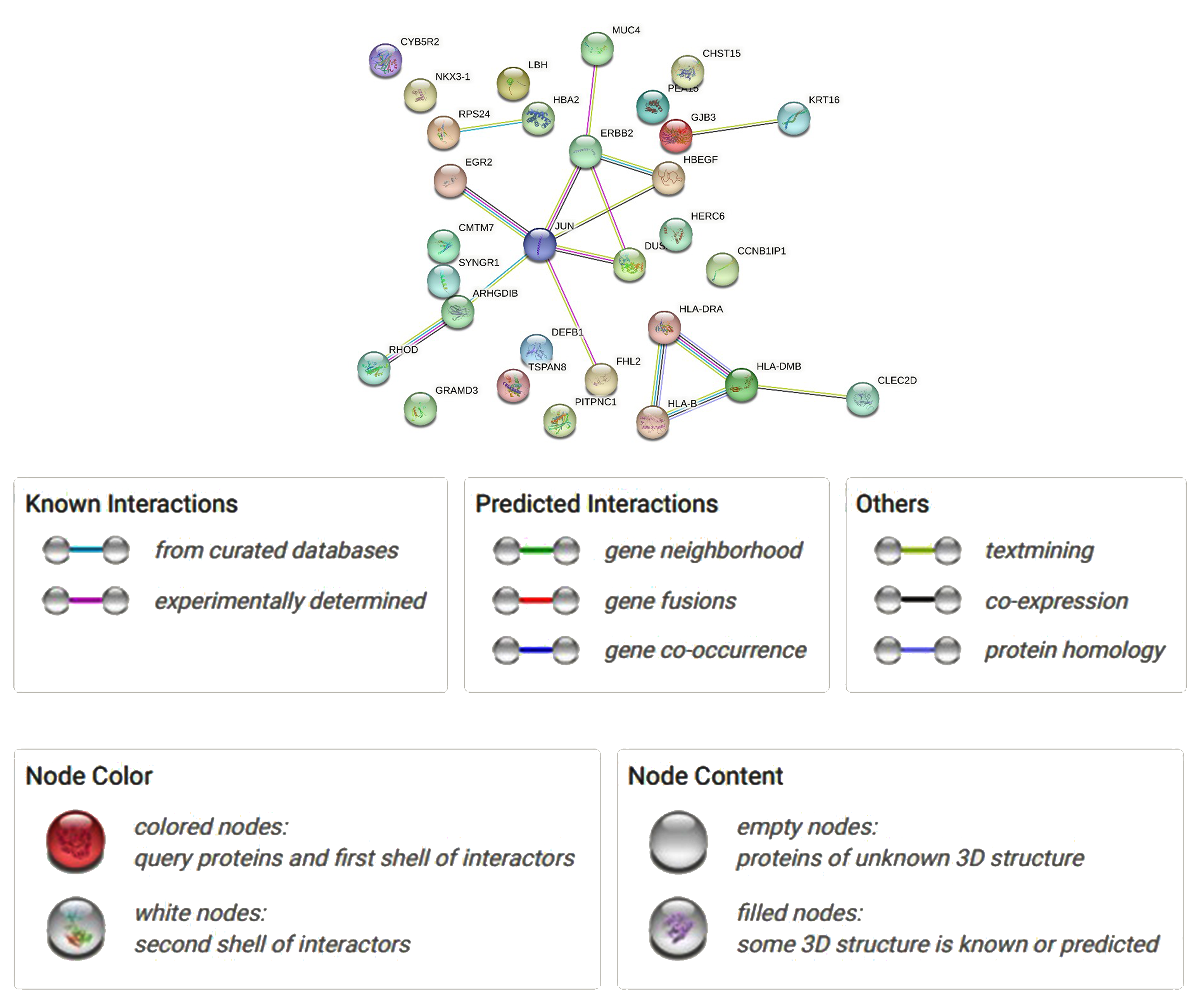}
    \caption{Protein-Protein interaction Network through String.}
    \label{fig: String PPI}
\end{figure}

\begin{figure}[!htbp]
    \centering
    \includegraphics[scale=.07]{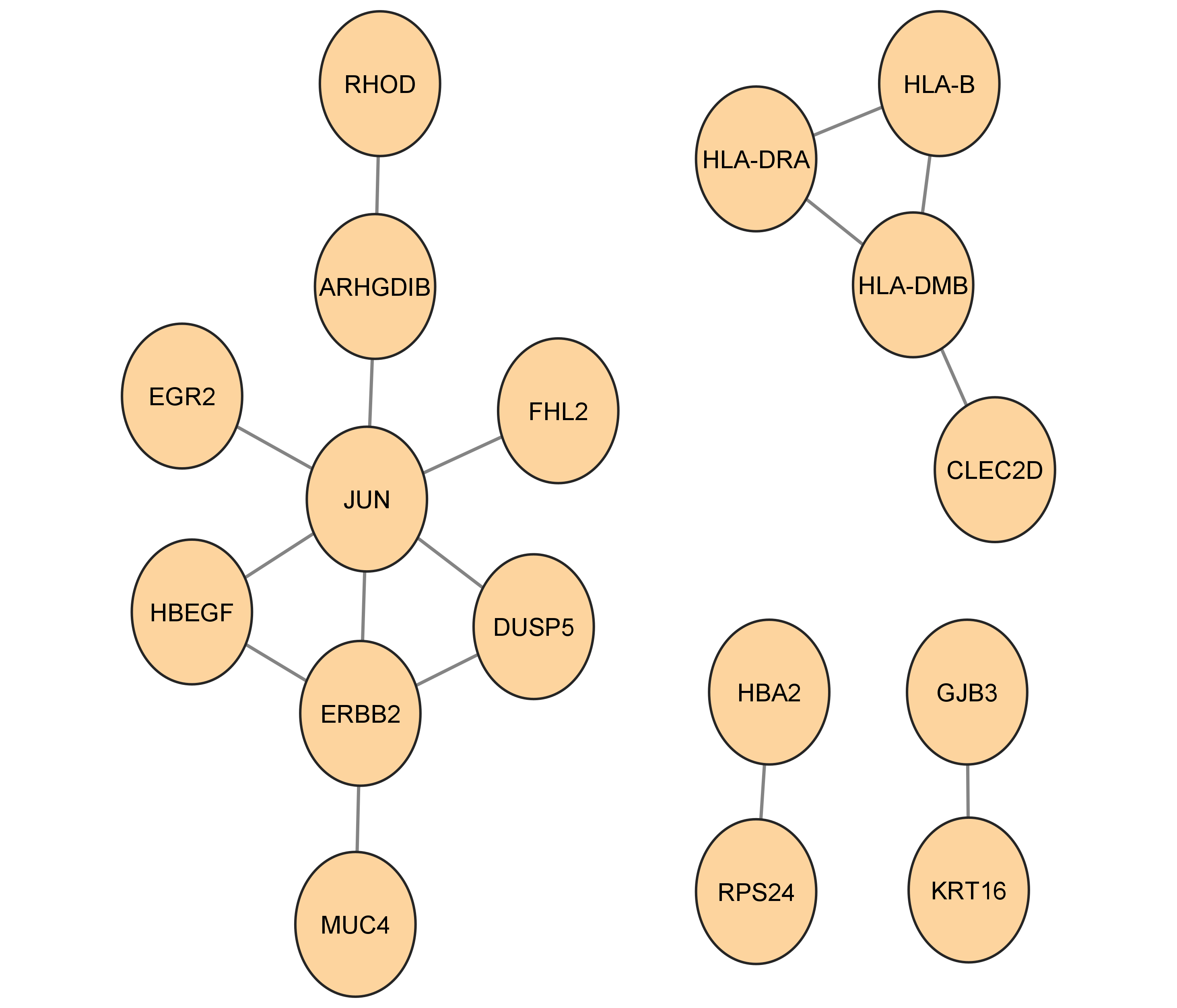}
    \caption{PPIs Network through Cytoscape using the directly interconnected genes.}
    \label{fig: Cytoscape PPI}
\end{figure}

\begin{figure}[!htbp]
    \centering
    \includegraphics[scale=.75]{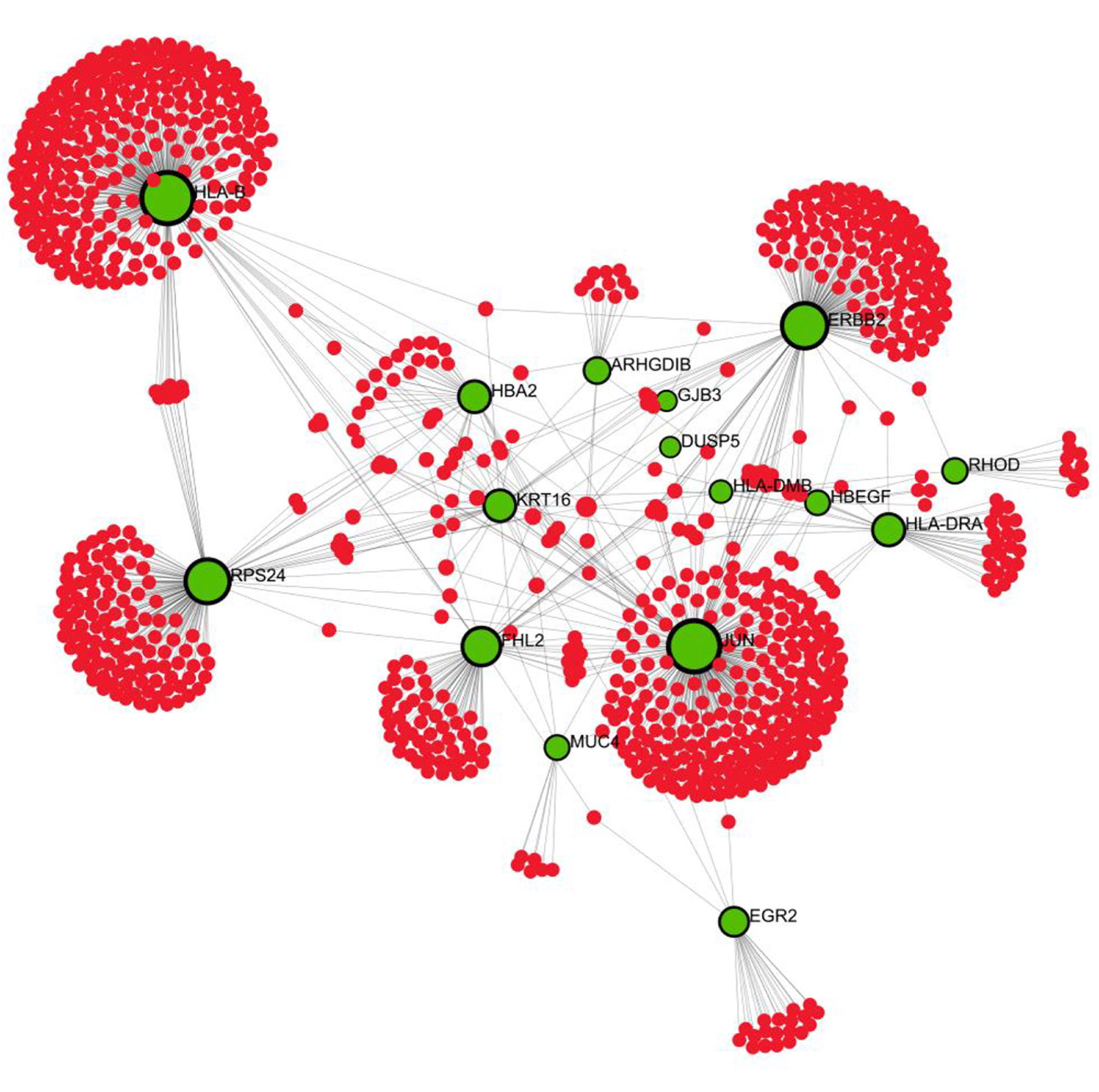}
    \caption{PPIs Network obtained through IMeX intercome of InnateDB Database contains 972 nodes, 1110 edges, and 16 seeds.}
    \label{fig: IMeX PPI}
\end{figure}

The network of protein-protein interactions was built using the STRING \ref {fig: String PPI}. Interconnected genes and disconnected genes are easily differentiated from this Figure \ref{fig: String PPI}. This Network was further evoked to Cytoscape for better visualization. This PPI network shown in Figure \ref{fig: Cytoscape PPI}, contains only 17 connected genes. Another PPI network was contrived by IMEx Interactome of NetworkAnalyst using the corresponding connected genes to understand the infection state by those corresponding genes. This network shown in Figure \ref{fig: IMeX PPI}, contains 972 nodes, 1110 edges, and 16 seeds.  These 16 seeds are those 17 interconnected genes except “CLEC2D”. These 16 seeds have a higher degree of interaction with the Protein. As “CLEC2D” had no significant interactions in the network. So, the IMEx Interactome automatically removed “CLEC2D” as the network seed. 

\subsection{Pinpointing Hub Genes}
\begin{figure}[!htbp]
    \centering
    \includegraphics[scale=.3]{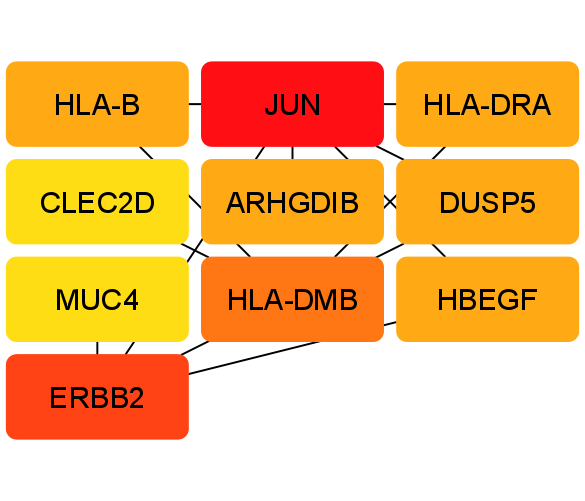}
    \caption{The Top 10 Hub genes Network according to the Degree Topology Method through Cytoscape.}
    \label{fig: Degree}
\end{figure}

\begin{figure}[!h]
    \centering
    \includegraphics[scale=.3]{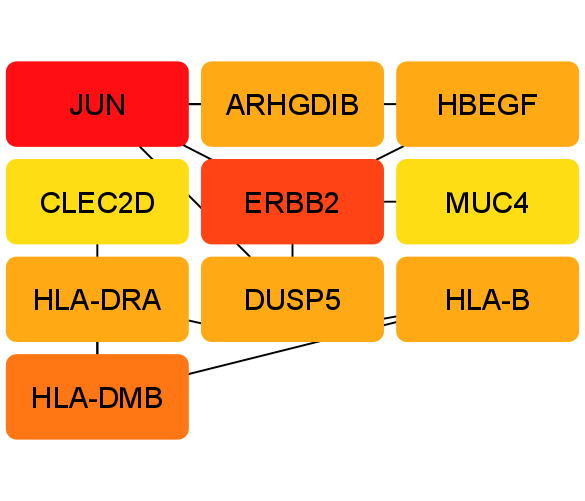}
    \caption{The Top 10 Hub genes Network accordance with The maximal clique centrality (MCC) Method through Cytoscape.}
    \label{fig: MCC}
\end{figure}

 For this research, the top 10 hub genes were taken. Because these top 10 hub genes are considered to be the most responsible genes among all. If these 10 hub genes are cured by a therapeutic molecule, then all other affected genes may have the possibility to recover as well as these hub genes have interaction with other genes.  10 hub genes had been pinpointed from the reconstructed PPI network as shown in Figure \ref{fig: Cytoscape PPI} of Cytoscape using the Degree topology and  the maximal clique centrality (MCC) method. Table \ref{table:HubGenes} and \ref{table:HubGenes1} show the top 10 hub genes according to the degree topology method and the maximal clique centrality (MCC) method respectively. The JUN has the highest interaction among the retrieved 10 hub genes.  The same hub genes were retrieved using 2 different methods, the Degree topology method and The maximal clique centrality (MCC) method. JUN and CLEC2D have the highest and lowest scores, respectively, in both methods. Seeing that from the TF-gene interactions network and the Gene-miRNA network, the hub genes with score 1 have contributed to those networks. As microRNAs (miRNAs) are discovered to promote mRNA degradation or prevent post-transcriptional translation understanding the functions of pleiotropic global regulators requires identifying the significant TF-gene interactions. After that 2 networks were drawn through The Cytoscape. The Grid Layout was chosen to extract the Figure \ref{fig: Degree} and the Figure \ref{fig: MCC}. 

\begin{table}[!h]
    \begin{center}
    \caption{The 10 Hub Genes, ordered by degree of importance}
        \label{table:HubGenes}
        \begin{tabular}{ |l|l| }
        \hline
            Name & Score \\
             \hline
             JUN & 6 \\
             \hline
             ERBB2 & 4 \\
             \hline
             HLA-DMB & 3 \\
             \hline
             HBEGF & 2 \\
             \hline
             HLA-B & 2 \\
		  \hline
             HLA-DRA & 2 \\
		\hline
             DUSP5 & 2 \\
		\hline
             ARHGDIB & 2 \\
		\hline
             MUC4 & 1 \\
		\hline
             CLEC2D & 1 \\
             \hline
        \end{tabular}
    \end{center}
\end{table}

\begin{table}[!h]
    \begin{center}
    \caption{The 10 Hub Genes, ordered by MCC (The maximal clique centrality) Method.}
        \label{table:HubGenes1}
        \begin{tabular}{ |l|l| }
        \hline
            Name & Score \\
             \hline
             JUN & 7 \\
             \hline
             ERBB2 & 5 \\
             \hline
             HLA-DMB & 3 \\
             \hline
             HBEGF & 2 \\
             \hline
             HLA-B & 2 \\
		  \hline
             HLA-DRA & 2 \\
		\hline
             DUSP5 & 2 \\
		\hline
             ARHGDIB & 2 \\
		\hline
             MUC4 & 1 \\
		\hline
             CLEC2D & 1 \\
             \hline
        \end{tabular}
    \end{center}
\end{table}

\subsection{Functional association Network}
\begin{figure}[!htbp]
    \centering
    \includegraphics[scale=.55]{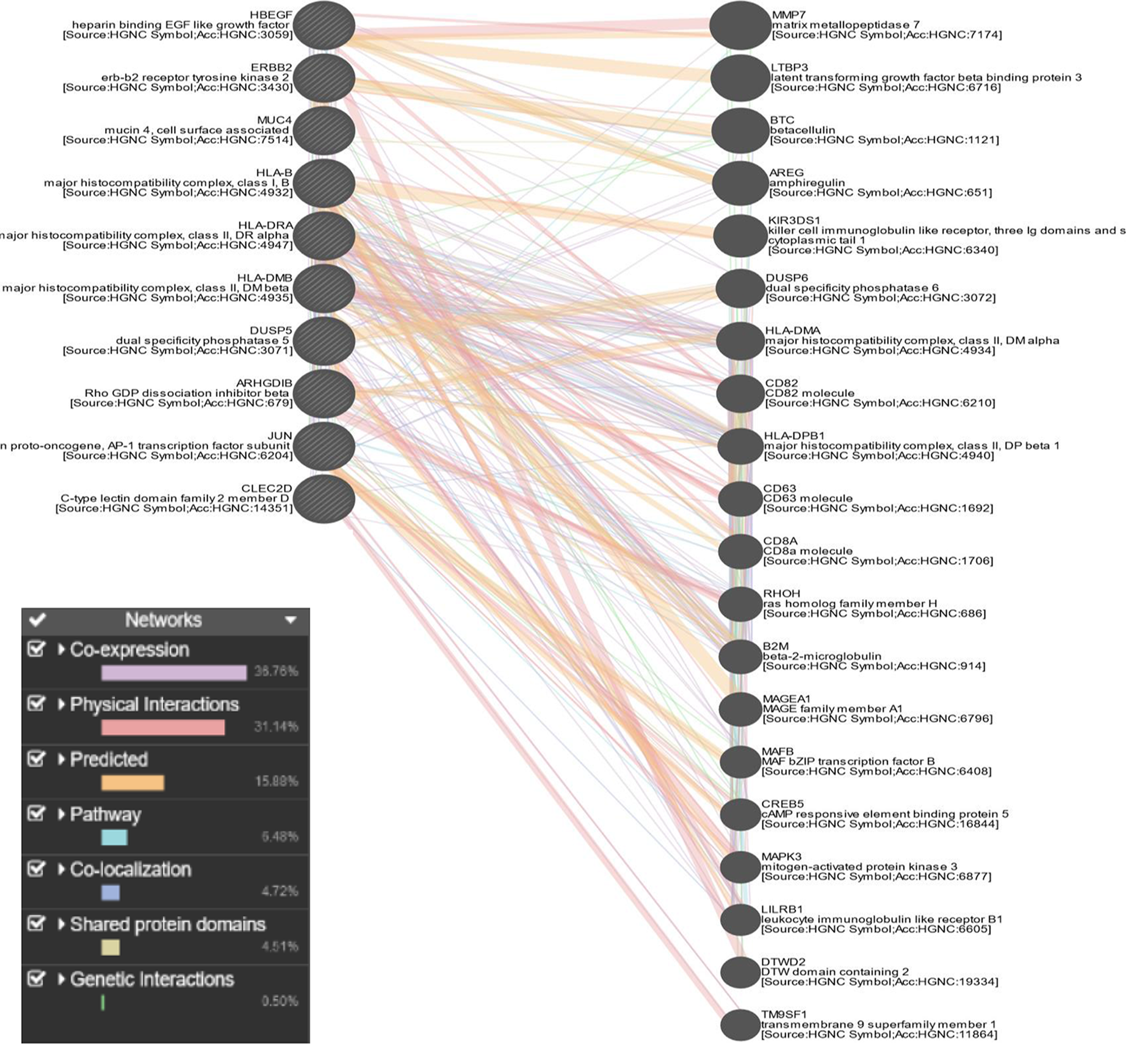}
    \caption{Functional association Network through GeneMania. 36.76\% Co-expression, 31.14\% Physical Interactions, 15.88\% Predicted, 6.48\% Pathway, 4.72\% Co-localization, 4.51\% Shared protein domains and 0.50\% Genetic Interactions are found here.}
    \label{fig: genemania}
\end{figure}

The 10 hub genes were used in the GeneMania Functional Network. This Figure \ref{fig: genemania} helped us to predict how certain gene sets will behave.  Utilizing a massive collection of functional association data, GeneMANIA discovers additional genes that are connected to a set of input genes. Protein and genetic relationships, pathways, co-expression, co-localization, and protein domain similarity are all examples of association data. GeneMANIA can be used to discover new components of a pathway or complex, discover extra genes that may have escaped existing screens, or discover novel genes that have a particular function, such as protein kinases. If the input gene list has five or more genes, GeneMANIA uses the "assigned based on query gene" technique to assign weights to enhance connectivity between all of the given input genes. To maximize the interaction between genes on a given list and minimize the interaction with genes not on a given list, the weights are automatically selected using linear regression. As our input gene list had more than 5 genes (10 Hub genes), this default method was done here.  This network has 30 nodes and displays functional keywords such as shared protein domains, co-expression, physical interactions, predicted, pathways, and genetic interactions \cite{warde2010genemania} and also shows the percentage of the functional keywords for our interested gene set.  36.76\% Co-expression, 31.14\% Physical Interactions, 15.88\% predicted, 6.48\% Pathway, 4.72\% Co-localization, 4.51\% Shared protein domains and 0.50\% Genetic Interactions are found in the Functional Association Network. 
A higher level of co-expression (36.76\%) between the transcripts associated with the two selected diseases and a consistent 31.14\% physical interaction. This implies a more pronounced connection at the genetic and molecular levels between the diseases. The acknowledgment of these strong associations is then used to support the idea that it is relatively straightforward to customize or modify a generic medication to effectively treat both distinct illnesses. The thesis here is that a single medicine can target a shared biological basis shared by the two diseases because of the significant co-expression and physical interaction between the genes and proteins linked to them. It might be easier to create a drug that treats both illnesses at once because of their similar foundation.

\subsection{Association of TF-gene}
\begin{figure}[!htbp]
    \centering
    \includegraphics[scale=.70]{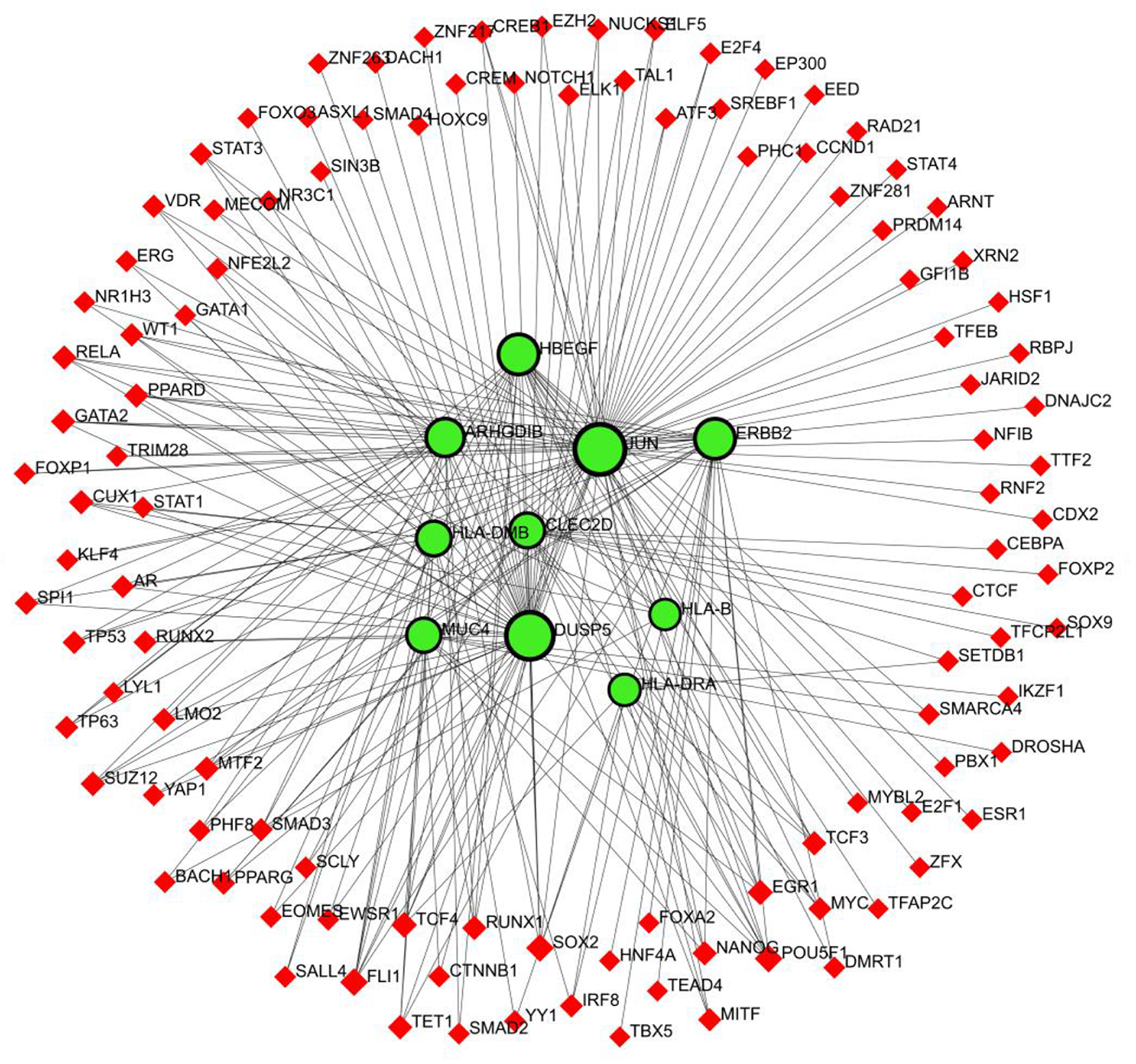}
    \caption{Visualization of TF-gene association network through NetworkAnalyst. Red Diamond Shaped Nodes indicate TF-gene and Green-Black highlighted circle-shaped nodes for seeds}
    \label{fig: TF-Gene}
\end{figure}

The TF-gene association network was fabricated to understand the transcription factor produced by the interested gene set. It was contrived using NetworkAnalyst. Hub genes (JUN, ERBB2, HLA-DMB, HBEGF, HLA-B, HLA-DRA, DUSP5, ARHGDIB, MUC4, CLEC2D) were used to build a TF-gene interaction network.  Ten hub genes were given as “Gene Input List” in NetworkAnalyst. Then “H.sapiens (human)” and “Official Gene Symbol” were chosen for “Specify organism” and “Set ID type” attributes correspondingly. By uploading that information, TF-gene interactions were selected from the Gene Regulatory Interaction. From the available 3 options (ENCODE, JASPAR, and ChEA) ChEA was selected to draw this network. ChEA database is a transcription factor targets database inferred from integrating literature curated Chip-X data. After proceeding further, the network is constructed via the ChEA database. Background as White and layout as Circular Bi/Tripartite were chosen to better visualize the network. Also, Opacity, Thickness, Color, Label, and size were customized from Edge and Node options. Red color for TF-gene and green with the black highlight for seeds were selected from The Global Node Styles for better visualization of this network. 
There are 119 nodes, 226 edges, and 10 seeds in the TF-Gene network as shown in Figure \ref{fig: TF-Gene}. \ colour {blue}  In our constructed TF-gene interaction, JUN is regulated by 60 TFs, The 47 TFs that control DUSP5, 24 TFs for HBEGF, and 24 TFs for ERBB2, and  HLA-DMB, HLA-B, HLA-DRA, ARHGDIB, MUC4, CLEC2D are regulated by 13 TFs, 7 TFs,8 TFs,19 TFs,12 TFs and 13 TFs respectively. 

\subsection{Gene-miRNA interactions}
\begin{figure}[!htbp]
    \centering
    \includegraphics[scale=.62]{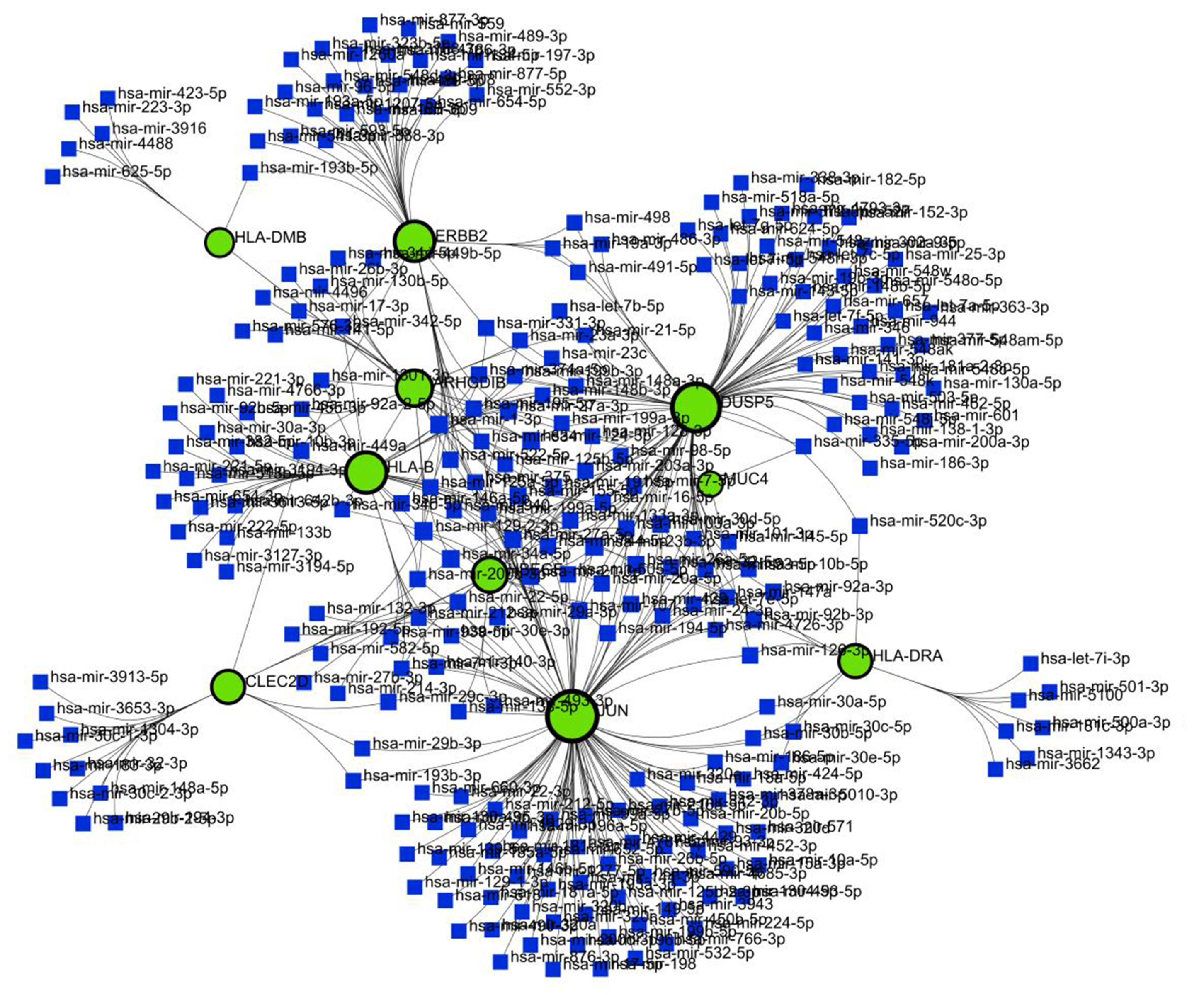}
    \caption{Visualization of Gene-miRNA network through NetworkAnalyst. Green-black highlighted Nodes indicate 17 seeds and Blue box-shaped nodes for miRNA, Edges connect the Genes and miRNAs}
    \label{fig: Gene-miRNA}
\end{figure}

Using NetworkAnalyst, the same Hub genes (JUN, ERBB2, HLA-DMB, HBEGF, HLA-B, HLA-DRA, DUSP5, ARHGDIB, MUC4, CLEC2D) were used as input to display the gene-miRNA interaction network. The network as shown in Figure \ref{fig: Gene-miRNA} comes with 271 nodes, 401 edges, and 10 seeds. The gene-miRNA interaction network shows miRNAs originated from those 10 hub genes. JUN is regulated by 118 miRNAs, The 96 miRNAs that control DUSP5, 43 miRNAs and 47 miRNAs respectively control ERBB2 and HLA-B.

\subsection{Interactions of Gene-disease}
\begin{figure}[!htbp]
    \centering
    \includegraphics[scale=.30]{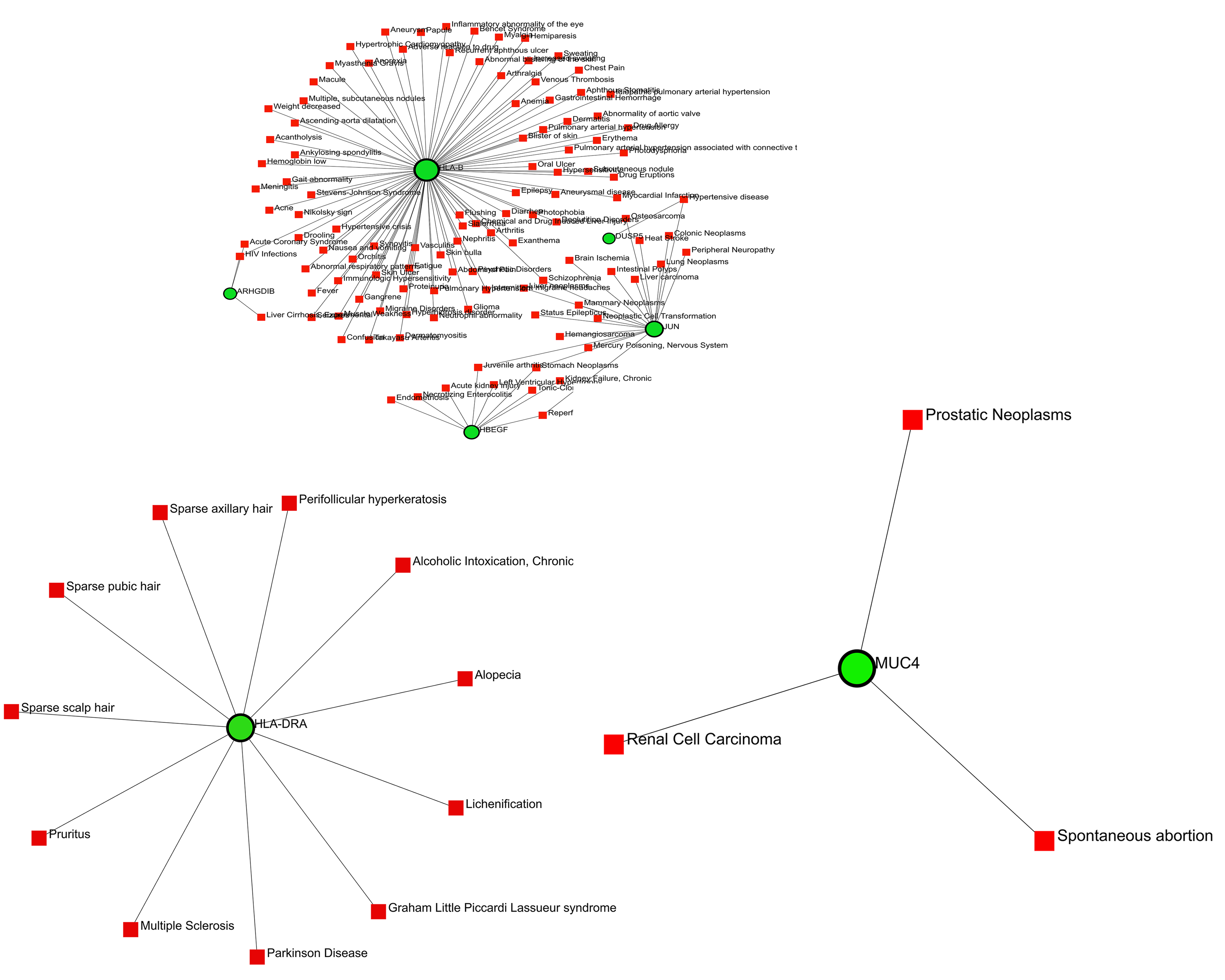}
    \caption{Gene-disease Network is divided into 3 subnetworks. Subnetwork1 represents genes (HLA-B, DUSP5, ARHGDIB, JUN, HBEGF) and their corresponding associated diseases, Subnetwork2 acts for HLA-DRA genes with its associated diseases and Subnetwork3 focuses on the MUC4 gene with its correspondent genes.  Here, Green-Black highlighted Nodes for seeds and red box-shaped nodes for associated diseases.}
    \label{fig: GENE Disease}
\end{figure}

DisGeNET database under NetworkAnalyst was used to demonstrate the Gene-disease association network. Hub genes (JUN, ERBB2, HLA-DMB, HBEGF, HLA-B, HLA-DRA, DUSP5, ARHGDIB, MUC4, CLEC2D) were also used to construct a Gene-disease association network.   DisGeNET database under NetworkAnalyst was used here to build the gene-disease correlations Network. 10 hub genes were given as “Gene Input List” in NetworkAnalyst. Then “H.sapiens (human)” and “Official Gene Symbol” were chosen for “Specify organism” and “Set ID type” attributes correspondingly. By uploading that information, Gene-disease associations were selected from Diseases, drugs \& chemicals. A gene-disease association database has information about gene-disease associations that have been curated by the literature gathered from the DisGeNET database that only applies to human data. After proceeding further, the network is constructed via the DisGeNET database.  The network contains 3 subnetworks.  Each subnetwork contains the genes with their associated disease.  This network is divided into 3 subnetworks because each subnetwork has no common associated diseases and contains its corresponding diseases individually. SubNetwork1 contains 117 nodes, 118 edges, and 5 seeds, and Subnetwork2 contains 12 nodes, 11 edges, 1 seed,  and Subnetwork3 contains 4 nodes, 3 edges, and 1 seed. The network as shown in Figure \ref{fig: GENE Disease} has 133 nodes, 132 edges, and 7 seeds in total.

\subsection{Common Drug suggestion}
Those 10 hub genes (JUN, ERBB2, HLA-DMB, HBEGF, HLA-B, HLA-DRA, DUSP5, ARHGDIB, MUC4, CLEC2D) by the degree topology method and The maximum clique centrality (MCC)  were used to recommend common drugs for the selected two diseases. These 10 genes are the most responsible genes of all.    From the Gene-disease association network, it is clear that the 10 hub genes are also responsible for some new diseases. That means, some new diseases will be occurred due to those affected genes. So these most affected genes must be cured by therapeutic molecules. 10 hub genes were used to recommend common drugs for the selected two diseases. These 10 hub genes are considered to be the most responsible among all genes. So these 10 hub genes must be cured by any molecules. When any molecule is used to cure these hub genes, it also affects other connected genes with those hub genes. 10 hub genes were given as input in the Enrichr platform. From the available options, the Diseases\/Drugs option was chosen. Our suggested drugs were retrieved from the DSigDB database under the Disease\/Drugs option in the Enrichr platform.  Table \ref{table: drugs} shows some predicated drug compounds for Hypopharyngeal cancer and EGFR-mutated lung adenocarcinoma patients who have these 2 diseases simultaneously. Here, 8 well-known therapeutic molecules were suggested from the DSigDB database. We suggest these 8 common drugs based on the number of hub genes cured.

\begin{table}[!htbp]
    \begin{center}
    \caption{EGFR-mutated lung adenocarcinoma patients and those with Hypopharyngeal cancer: predicated Drugs}
        \label{table: drugs}
        \begin{tabular}{ |l|l|p{4cm}| }
        \hline
           Drug's Name & Adjusted P-value & Genes\\
             \hline
             Retinoic acid  CTD 00006918 & 0.007990864 & DUSP5;JUN;ERBB2; ARHGDIB;HLA-B;HLA-DRA;MUC4;HBEGF \\
             \hline
             Arsenenous acid CTD 00000922 & 0.010633891 & JUN;ERBB2;ARHGDIB; HLA-B;HBEGF\\
	           \hline
             TERT-BUTYL HYDROPEROXIDE CTD 00007349 & 0.012283126 & DUSP5;JUN;ERBB2; HLA-B;HBEGF\\
             \hline
             carbamazepine CTD 00005574 & 0.007027547 & JUN;ERBB2;HLA-B;HBEGF \\
             \hline
               etoposide CTD 00005948 & 0.007027547 & JUN;ERBB2;ARHGDIB; HBEGF\\
	           \hline
            tonzonium bromide PC3 UP & 0.007027547 & DUSP5;JUN;HBEGF \\
             \hline
             NVP-TAE684 CTD 00004657 & 0.007027547 & ERBB2;HBEGF \\
             \hline
             prostaglandin J2 CTD 00001744 & 0.007027547 & JUN;ARHGDIB \\
             \hline
             \end{tabular}
    \end{center}
\end{table}

\section{Discussion}
A dangerous constituent for Hypopharyngeal cancer is lung adenocarcinoma with EGFR mutation, which are reported to have some associations with each other. Therefore, we hypothesized that this association is due to some common genes/proteins as drivers. To do the investigation, microarray datasets are collected accordingly, followed by differential expression analysis with the Benjamini-Hochberg False Discovery rate (FDR) approach.  Common DEGs from the two datasets were then gathered once the differentially expressed genes had been found from the dataset. The threshold for FDR was chosen as $<0.05$ as it is quite common in statistical hypothesis testing relevant literature. While we use a stringent FDR  ($<0.05$)  cut-off to retrieve DEGs, a substantially large number of common genes are retrieved.  If we apply stringent FDR (i.e., $<0.10$), the necessary (false positive) genes have been  also chanced to deduct. If important genes are mistakenly excluded from the analysis, it can lead to inaccurate identification of differentially expressed genes between conditions. These deducted genes can also affect the entire downstream analysis pipeline with false positive results. This may transduce to erroneous drug prediction, which was very crucial for our aim, as we wanted to find out the targeted molecules for 2 different diseases. So, we have to find out the most effective responsible genes rather than retrieving a large number of common genes. 32 genes (GRAMD3, SYNGR1, CMTM7, HLA-DMB, HLA-DRB4, CLEC2D, CCNB1IP1, HLA-DRA, DEFB1, ERBB2, MUC4, LOC145837, RPS24, RHOD, HBEGF, ARHGDIB, PITPNC1, PEA15, KRT16, GJB3, JUN, FHL2, CYB5R2, HLA-B, EGR2, HERC6, DUSP5, HBA2, CHST15, NKX3-1, LBH, TSPAN8) are common in Hypopharyngeal cancer and EGFR-mutated lung adenocarcinoma. Further analysis was done by using these common DEGs.

To discover GO terms and pathways, an analysis of gene set enrichment was performed. The PPI network is the most notable part for detecting interconnected genes, disconnected genes, and hub genes identification. HLA-B, JUN, ERBB2, RPS24, FHL2, HLA-DRA, KRT16, HBA2, EGR2, ARHGDIB, HLA-DMB, CLEC2D, MUC4, RHOD, HBEGF, GJB3, DUSP5 were pinpointed as directly interconnected genes and JUN, ERBB2, HLA-DMB, HBEGF, HLA-B, HLA-DRA, DUSP5, ARHGDIB, MUC4, CLEC2D were spotted as Hub genes according to degree method and The maximum clique centrality (MCC). 

MicorRNAs are gene silencing factor. Common genes were used for retrieving their corresponding miRNAs and TFs. 32 miRNAs and 33 TFs were produced for the targeted gene set. The network has 93 nodes, 223 edges, and 28 seeds. The miRNA has-miR-27a has 5 which is the highest interaction among all the miRNAs and the MAX TF gene has 8 which is also the highest interaction among all TFs in TF-miRNA coregulatory network. 

TF-gene network was contrived by 10 hub genes of Hypopharyngeal cancer and EGFR-mutated lung adenocarcinoma. TFs are the promoter genes that are responsible for transcription. The control of gene expression is carried out by particular genes that interact with TF genes, which act as reactors for this control \cite{zhang2015transcription}. JUN has an elevated interconnection among all in the network in our study. Another gene-miRNA network was analyzed for those hub genes that were identified earlier in the PPI network. miRNAs that can regulate gene expression by slowing down mRNA synthesis \cite{zhang2015transcription}. To know the miRNAs produced for those 10 hub genes, a gene-miRNA network was fabricated. In our investigation, 261 miRNAs are produced for those 10 hub genes. The gene-disease association network was constructed for the same hub genes. To know the risk genes among them, which can be cause for other associated diseases. If these risk genes can't be cured by any molecules then these associated diseases can occur in the near future.

Note, in this study, we have collected the Protein-Protein Interaction information from the STRING database. In this database, PPI information is defined based on various types of gene/protein association evidence, e.g., known interactions (curated databases, experimentally verified interactions), predicted interactions (gene neighborhood, gene fusion, gene co-occurrences), and other types of associations (text-mining, coexpression, protein homology). Therefore, we argue that the PPI information used in our study is substantially comprehensive, i.e., covering not only protein-protein association but also functional association (indirectly), and TF-gene interaction. However, for Gene-miRNA and Gene-disease interaction, the STRING database is no evidence of such types evidence. However, for our future works, we aim to augment those types of gene/protein associations (Gene-miRNA and Gene-disease) along with STRING PPI, to form an integrated network, upon which we will conduct further downstream analysis.

Finally, drug molecules based on the 10 hub genes were recommended. These 10 genes are considered the most affected and responsible genes among all. If we cure these genes with any therapeutic molecules, other genes that are connected to these 10 genes will be also cured by the same therapeutic molecules. If these 10 genes are not cured by molecules, other associated diseases can occur due to these genes. From our suggested 8 drugs, the Retinoic acid CTD 00006918 can affect 8 hub genes among the 10 hub genes. Patients having both diseases (Hypopharyngeal cancer and EGFR-mutated lung adenocarcinoma) concurrently may have a higher possibility of cure by using our suggested drug compounds. Our suggested drugs may be approved in the future after doing further chemical experiments, testing, and so on. 
If some related illnesses are found with Hypopharyngeal Cancer and EGFR-mutated lung adenocarcinoma then future research in this area aims to create a single generic drug to treat some related illnesses, offering a fresh perspective.

\section{Conclusion}
\label{conclusion}
This study mentioned that the selected two diseases may have the possibility to metastasize to one another. Analyzing any disease means analyzing the disease genes. The constructed PPI Networks displayed all the Directly associated genes, general genes, and a channel that ensures the route to a general remedy map. The Cytohubba module was used to identify 10 hub genes using the Degree Topology approach and the maximum clique centrality (MCC).  Only those genes that are interconnected with each other and have higher interaction among all are taken for this research purpose. If we can recover the directly affected, higher interconnected genes of a disease, we can get rid of those selected diseases (Hypopharyngeal cancer and EGFR-mutated Lung Adenocarcinoma). The next step is to employ GeneMania to develop a new network for the 32 shared genes to learn more about their physical interactions, shared protein domains, shared pathways, and genetic interactions. TF-gene, Gene-miRNA, and Gene-Disease association networks were designed by using the same 10 hub genes. After analyzing those networks, some well-known therapeutic molecules were suggested for Hypopharyngeal cancer and EGFR-mutated lung adenocarcinoma by using the 10 hub genes as input. A common drug for selected two associated diseases aims to reduce the amount of drug one should take and also reduce cost. Future research in this area aims to create a single generic drug to treat several related illnesses, offering a fresh perspective.

\section*{Declarations}

\subsection*{Conflict of interest}
The authors have no conflicts of interest to declare that they are relevant to the content of this article.

\section*{Acknowledgments}
This work is supported by the Special Research Grant, ICT Division, Bangladesh, 2022-2023 "Machine Learning and Bioinformatics for disease genes identifications, biomarker \& drug discovery".

\subsection*{Ethics approval}  Not applicable
\subsection*{Consent to participate} Not applicable
\subsection*{Consent to Publish} Not applicable

\subsection*{Availability of data and materials}

The selected datasets are sourced from free and open-access sources such as
The GEO database \url{https://www.ncbi.nlm.nih.gov/geo/}.

\subsection*{Authors’ contributions}
Abanti Bhattacharjya: Data curation, Methodology, Software, Writing- Original draft preparation; Md. Manowarul Islam: Conceptualization, Supervision, Methodology, Formal analysis, Visualization, Writing- Original draft preparation; Md Ashraf Uddin: Visualization, Investigation, Formal analysis; Md. Alamin Talukder: Investigation, Validation, Visualization, Writing - Reviewing and Editing; AKM Azad: Visualization, Validation, Writing- Reviewing and Editing; Sunil Aryal: Investigation, Validation; Bikash Kumar Paul and Wahia Tasnim: Visualization; Muhammad Ali Abdulllah Almoyad: Validation and Mohammad Ali Moni: Analysis, Validation, Visualization, Writing- Reviewing and Editing.

\bibliography{bibs}

\end{document}